\documentclass[letterpaper,12pt]{article}

\usepackage{times}
\usepackage{amsthm}
\usepackage{amssymb} 
\usepackage{booktabs} 
\usepackage{outlines}
\usepackage{xspace}
\usepackage{amsmath}
\usepackage{mathtools}
\usepackage{subcaption}
\usepackage{enumitem}
\usepackage{graphicx}
\usepackage{nicefrac}
\usepackage{listings}
\usepackage{color}
\PassOptionsToPackage{hyphens}{url}\usepackage{hyperref}
\usepackage{ragged2e}
\usepackage{authblk}
\usepackage[labelfont=bf]{caption}
\usepackage[margin=1in,letterpaper]{geometry} 
\usepackage{url}

\usepackage{hyperref}
\usepackage{array}   
\newcolumntype{L}{>{$}l<{$}} 

\newtheorem{prop}{Proposition}

\usepackage{titlesec}
\usepackage{eurosym}

\begin{document}




\title{\textbf{\LARGE High Throughput Cryptocurrency Routing in Payment Channel Networks}}

\author[1]{Vibhaalakshmi Sivaraman}
\author[2]{Shaileshh Bojja Venkatakrishnan}
\author[3]{Kathleen Ruan}
\author[1]{Parimarjan Negi}
\author[1]{Lei Yang}
\author[4]{Radhika Mittal}
\author[1]{\\ Mohammad Alizadeh}
\affil[1]{Massachusetts Institute of Technology}
\author[3]{Giulia Fanti}
\affil[2]{Ohio State University}
\affil[3]{Carnegie Mellon University}
\affil[3]{University of Illinois at Urbana-Champaign}

\date{}

\newcommand{\ignore}[1]{}
\newcommand{\TODO}[1]{\ifdraft {\color{red} \textbf{TODO:} {#1}} \fi }

\newcommand{\vls}[1]{{\color{blue} \textbf{Vibhaa:} {#1}}}
\newcommand{\sbv}[1]{{\color{green} \textbf{Shaileshh:} {#1}}}
\newcommand{\gf}[1]{{\color{red} \textbf{Giulia:} {#1}}}

\newcommand{\SM}{SpeedyMurmurs\xspace}
\newcommand{\SW}{SilentWhispers\xspace}
\newcommand{\MF}{Max-flow\xspace}
\newcommand{\WF}{Waterfilling}
\newcommand{\LR}{Landmark Routing\xspace}
\newcommand{\tu}{transaction-unit\xspace}
\newcommand{\eu}{\euro{}\xspace}
\newcommand{\tus}{transaction-units\xspace}
\newcommand{\NT}{normalized throughput\xspace}
\newcommand{\SV}{success volume\xspace}
\newcommand{\SR}{success ratio\xspace}
\newcommand{\eg}{{\em e.g., }}
\newcommand{\etal}{{\em et al.}}
\newcommand{\etc}{{\em etc.}}
\newcommand{\vs}{{\em vs. }}
\newcommand{\ie}{{\em i.e., }}
\newcommand{\Sec}[1]{\S\ref{#1}}
\newcommand{\App}[1]{App.~\ref{#1}}
\newcommand{\TheSystem}{Spider\xspace}
\newcommand{\PCN}{PCN\xspace}
\newcommand{\Fig}[1]{Fig.~\ref{fig:#1}}
\newcommand{\NewPara}[1]{\noindent{\bf #1}}
\newcommand{\NewParaI}[1]{\noindent{\emph #1}}
\newcommand\blfootnote[1]{%
  \begingroup
  \renewcommand\thefootnote{}\footnote{#1}%
  \addtocounter{footnote}{-1}%
  \endgroup
}
\newcommand{\eat}[1]{}
\newcommand{\Tab}[1]{Tab.~\ref{tab:#1}\xspace}
\newcommand{\Eqn}[1]{Equation~\ref{eq:#1}\xspace}
\newcommand{\name}{Spider\xspace}

\maketitle

\begin{abstract}
Despite growing adoption of cryptocurrencies, making fast payments at scale remains a challenge. 
Payment channel networks (PCNs) such as the Lightning Network have emerged as a viable scaling solution.
However, completing payments on PCNs is challenging: payments must be routed on paths with sufficient funds.
As payments flow over a single channel (link) in the same direction, the channel eventually becomes depleted and cannot support further payments in that direction; hence, naive routing schemes like shortest-path routing can deplete key payment channels and paralyze the system. 
Today's PCNs also route payments atomically, worsening the problem.
In this paper, we present \name, a routing solution that ``packetizes'' transactions and uses
a multi-path transport protocol to achieve high-throughput routing in PCNs.
Packetization allows \name to complete even large transactions on low-capacity payment channels over time, 
while the multi-path congestion control protocol ensures balanced utilization of channels and fairness across flows.
Extensive simulations comparing \name with state-of-the-art approaches shows that \name requires less than 25\% of the funds to successfully route over 95\% of transactions on balanced traffic demands, and 
    offloads 4x more transactions onto the PCN on imbalanced demands.

\blfootnote{The lead author can be contacted at vibhaa@mit.edu}
\end{abstract}

\section{Introduction}
\label{sec:intro}

Despite their growing adoption, cryptocurrencies suffer from poor scalability. 
For example, the Bitcoin~\cite{bitcoin} network processes 7 transactions per second, and Ethereum~\cite{ethereum} 15 transactions/second, which pales in comparison to the 1,700 transactions per second achieved by the VISA network \cite{visa}. Scalability thus remains a major hurdle to the adoption of cryptocurrencies for retail and other large-scale applications.
The root of the scalability challenge is the inefficiency of the underlying consensus protocol: every transaction must go through full consensus to be confirmed, which can take anywhere from several minutes to hours~\cite{scale}.


%

A leading proposal among many  solutions to improve cryptocurrency scalability \cite{prism, algorand, omniledger} relies on so-called {\em payment channels}. 
A payment channel is a cryptocurrency transaction that escrows or 
dedicates money on the blockchain for exchange with a prespecified user for a predetermined duration.
For example, Alice can set up a payment channel with Bob in which she escrows 10 tokens for a month. 
Now Alice can send Bob (and only Bob) signed transactions from the escrow account, and Bob can validate them privately in a secure manner 
without mediation on the blockchain (\S\ref{sec:background}).
If Bob or Alice want to close the payment channel at any point, they can broadcast the most recent signed transaction message to the blockchain to finalize the transfer of funds.


The versatility of payment channels stems from \textbf{payment channel networks} (PCNs), in which users who do not share direct payment channels can route transactions through intermediaries for a nominal fee. PCNs enable fast, secure transactions without requiring consensus on the blockchain for every transaction.
PCNs have received a great deal of attention in recent years, and many blockchains are looking to PCNs to scale throughput
without overhauling the underlying consensus protocol.
For example, Bitcoin has deployed the Lightning network \cite{lightning-implementation,c-lightning-impl}, and Ethereum uses Raiden \cite{raiden}.

For PCNs to be economically viable, the network must be able to support high \emph{transaction throughput}.
This is necessary for intermediary nodes (routers) to profitably offset the opportunity cost of escrowing funds in payment channels, and for encouraging end-user adoption by providing an appealing quality of payment service.
But, a transaction is successful only if all channels along its route have sufficient funds. 
This makes payment channel {\em routing}, the protocol by which a path is chosen for a transaction, of paramount importance.


Existing payment channel routing protocols achieve poor throughput, for two main reasons. First, they attempt to route each incoming transaction atomically and instantaneously, in full.
This approach is harmful, particularly for larger transactions, because a transaction fails completely if there is no path to the destination with enough funds. Second, existing routing protocols fail to keep payment channels {\em balanced}. A payment channel becomes imbalanced when the transaction rate across it is higher in one direction than the other; the party making more transactions eventually runs out of funds and cannot send further payments without ``refilling'' the channel via either an on-chain transaction (i.e., committing a new transaction to the blockchain) or coordinated cyclic payments between a series of PCN nodes \cite{revive}. Most PCNs today route transactions naively on shortest paths with no consideration for channel balance; this can leave many channels depleted, reducing throughput for everyone in the network. We describe a third problem, the creation of {\em deadlocks} in certain scenarios, in \S\ref{sec:motivation}.

\eat{
Existing payment channel routing protocols achieve poor throughput for two main reasons. First, they fail to keep payment channels {\em balanced}. A payment channel becomes imbalanced when the transaction rate across it is higher in one direction than the other; the party making more transactions eventually runs out of funds and cannot send further payments without ``refilling'' the channel via an on-chain transaction (i.e., committing a new transaction to the blockchain). Most existing protocols route transactions naively on shortest paths. Since these path choices do not consider channel balance, they can leave many channels depleted, reducing throughput for everyone in the network. 
Second, these protocols route each incoming transaction atomically and instantaneously, in full.
This approach is particularly harmful for larger transactions, which can fail completely if there is no path to the destination with enough balance.
We describe a third problem, the creation of {\em deadlocks} in certain scenarios, in \S\ref{sec:motivation}.}

\eat{
The Lightning network is currently in its early days, and most of its transactions are still small \cite{small_tx}. 
For this reason, some of the issues we pointed out have not become critical yet.
However, as PCNs grow in adoption, transaction sizes are likely to become larger, and the distribution more
heavy-tailed.
Because of this, the listed problems could create serious scalability bottlenecks. 
}

In this paper we present {\em \name}, a multi-path transport protocol that achieves balanced, high-throughput routing in PCNs, building on concepts in an earlier position paper~\cite{spiderhotnets}. \name's design centers on two ideas that distinguish it from existing approaches. First, \name senders ``packetize'' transactions, splitting them into \tus that can be sent across different paths at different rates. By enabling congestion-control-like mechanisms for PCNs, this packet-switched approach makes it possible to send large payments on low-capacity payment channels over a period of time. Second, \name develops a simple multi-path congestion control algorithm that promotes balanced channels while maximizing throughput. \name's senders use a simple one-bit congestion signal from the routers to adjust window sizes, or the number of outstanding \tus, on each of their paths.  

\name's congestion control algorithm is similar to multi-path congestion control protocols like MPTCP~\cite{MPTCP} developed for Internet congestion control. But the routing problem it solves in PCNs differs from standard networks in crucial ways. Payment channels can only route transactions by moving a finite amount of funds from one end of the channel to the other. 
Because of this, the capacity of a payment channel\,---\,the transaction rate that it can support\,---\,varies depending on how it is used; a channel with balanced demand for routing transactions in both directions can support a higher rate than an imbalanced one.
Surprisingly, we find that a simple congestion control protocol can achieve such balanced routing, despite not being designed for that purpose explicitly.

We make the following contributions:
\begin{enumerate}[noitemsep,topsep=0pt,parsep=0pt,partopsep=0pt]
    \item We articulate challenges for high-throughput routing in payment channel networks (\S\ref{sec:motivation}),  and we formalize the balanced routing problem (\S\ref{sec:model}). We show that the maximum throughput achievable in a PCN depends
        on the nature of the transaction pattern: circulation demands (participants send on average as much
        as they receive) can be routed entirely with sufficient network capacity, while demands that form Directed Acyclic Graphs~(DAGs) where some
        participants send more than they receive cannot be routed entirely in a balanced manner. We also show that introducing
        DAG demands can create deadlocks that stall all payments.
    \item We propose a packet-switched architecture for PCNs (\S\ref{sec:architecture}) that splits transactions into \tus and multiplexes
        them across paths and time.
    \item We design \name (\S\ref{sec:protocol}), a multi-path transport protocol that (i) maintains balanced channels in the PCN, (ii) uses the funds escrowed in a PCN
        efficiently to achieve high throughput, and (iii) is fair to different payments.
    \item We build a packet-level simulator for PCNs and validate it with a small-scale implementation of \name on the LND Lightning Network codebase~\cite{lightning-implementation}. Our evaluations (\S\ref{sec:eval}) show that
    (i) on circulation demands where 100\% throughput is achievable, compared to the state-of-the-art, 
    \name requires 25\% of the funds to route over 95\% of the transactions 
    and completes 1.3-1.8x more of the largest 25\% of transactions based on a credit card transactions dataset~\cite{kaggledata};
    (ii) on DAG demands where 100\% throughput is not achievable, \name offloads
        7-8x as many transactions onto the PCN for every 
        transaction on the blockchain, a 4x improvement over current approaches.
\end{enumerate}



\section{Background}
\label{sec:background}

\begin{figure}[htbp]
\centering
\includegraphics[width=5in]{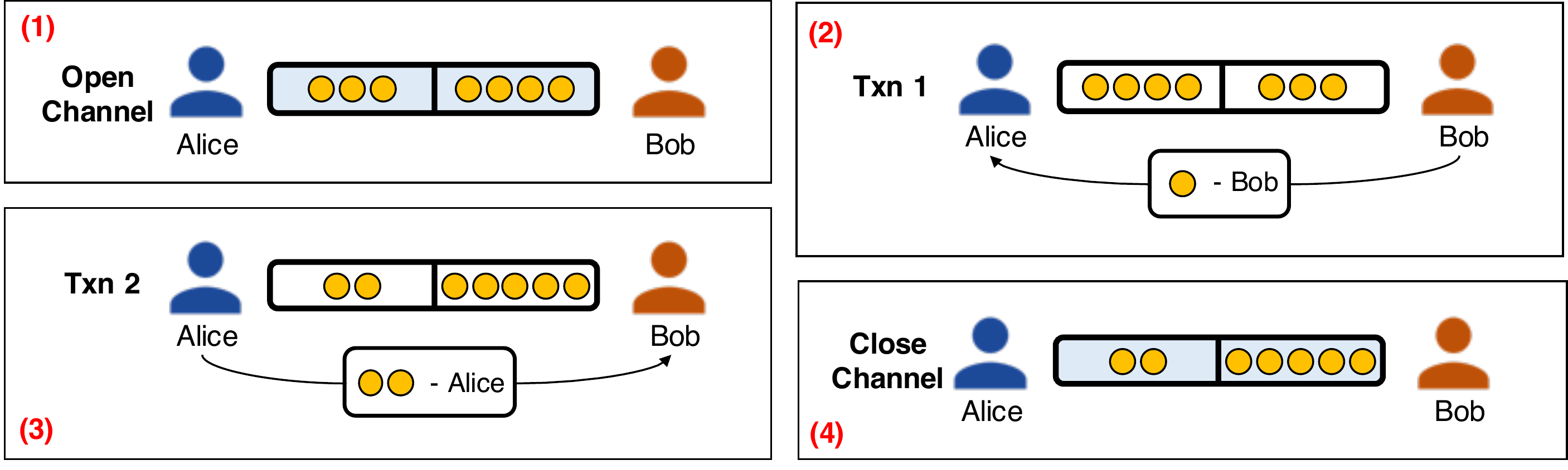}
\caption{\small Bidirectional payment channel between Alice and Bob. A blue shaded block indicates a transaction that is committed to the blockchain.}
\label{fig:pchannel}
\end{figure}

Bidirectional payment channels are the building blocks of a payment channel network. 
A bidirectional payment channel allows a sender (Alice) to send funds to a receiver (Bob) and vice versa. 
To open a payment channel, Alice and Bob jointly create a transaction that escrows money for a fixed amount of time \cite{lightning}.
Suppose Alice puts 3 units in the channel, and Bob puts 4 (\Fig{pchannel}). 
Now, if Bob wants to transfer one token to Alice, he sends her a cryptographically-signed message asserting that he approves the new balance. 
This message is not committed to the blockchain; Alice simply holds on to it. 
Later, if Alice wants to send two tokens to Bob, she sends a signed message to Bob approving the new balance (bottom left, \Fig{pchannel}). 
This continues until one  party decides to close the channel, at which point they publish the latest message to the blockchain asserting the channel balance. 
If one party tries to cheat by publishing an earlier balance, the cheating party loses all the money they escrowed to 
the other party \cite{lightning}. 

\begin{figure}[htbp]
\centering
\includegraphics[width=3in]{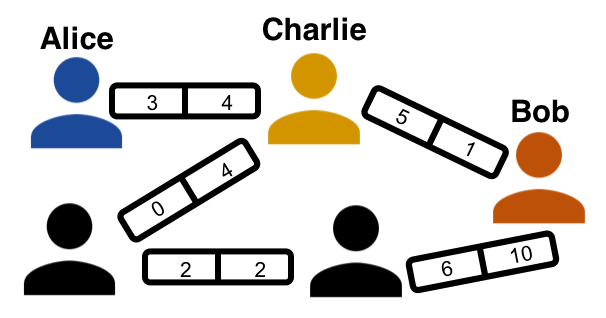}
\caption{\small In a payment channel network, Alice can transfer money to Bob by using intermediate nodes' channels as relays. There are two paths from Alice to Bob, but 
only the path (Alice, Charlie, Bob) can support 3 tokens.}
\label{fig:pcn}
\end{figure}



A payment channel network is a collection of bidirectional payment channels (\Fig{pcn}).
If Alice wants to send three tokens to Bob, she first finds a path to Bob that can support three tokens of payment. 
Intermediate nodes on the path (Charlie) will relay payments to their destination. 
Hence in \Fig{pcn}, two transactions occur: Alice to Charlie, and Charlie to Bob. 
To incentivize Charlie to participate, he receives a routing fee.
To prevent him from stealing funds, a cryptographic hash lock ensures that all intermediate transactions are only valid after a transaction recipient knows a private key generated by Alice \cite{raiden}.
\footnote{The protocol called Hashed Timelock Contracts (HTLCs) can be implemented in two ways: the sender generates the key, as
in Raiden \cite{raiden} or the receiver generates the key, as in Lightning \cite{lightning}. \name assumes that the
sender generates the key.}
Once Alice is ready to pay, she gives that key to Bob out-of-band; he can either broadcast it (if he decides to close the channel) or pass it to Charlie.
Charlie is incentivized to relay the key upstream to Alice so that he can also get paid. Note that Charlie's payment channels
with Alice and Bob are independent: Charlie cannot move funds between them without going through the blockchain.

\begin{figure*}[t]
    \centering
    \begin{tabular}{l c c}
        \begin{subfigure}{0.3\columnwidth}
            \centering
            \includegraphics[width=\columnwidth]{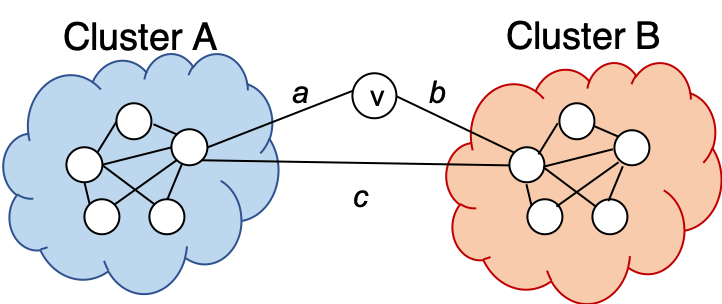}
            \subcaption{Underutilized channels}
\label{fig:barbell1}
        \end{subfigure} &
        \begin{subfigure}{0.3\columnwidth}
            \centering
            \includegraphics[width=\columnwidth]{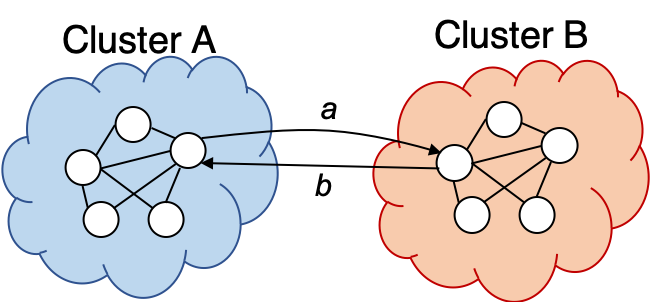}
            \subcaption{Imbalanced channels}
\label{fig:barbell2}
        \end{subfigure} &
                \begin{subfigure}{0.3\columnwidth}
            \centering
            \includegraphics[width=\columnwidth]{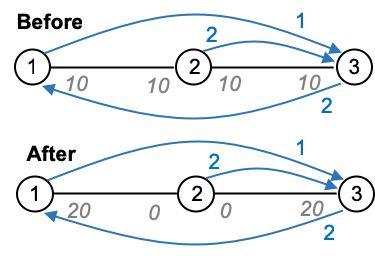}
            \subcaption{Deadlock}
\label{fig:deadlock}
        \end{subfigure} 
    \end{tabular}
\caption{\small Example illustrating the problems with state-of-the-art PCN routing schemes.}
\label{fig:motivation}
\end{figure*}

\section{Challenges in Payment Channel Networks}
\label{sec:motivation}

A major cost of running PCNs is the collateral needed to set up payment channels. 
As long as a channel is open, that collateral is locked up, incurring an opportunity cost for the owner. 
For PCNs to be financially viable, this opportunity cost should be offset by routing fees, which are charged on each transaction that passes through a router. 
To collect more routing fees, routers try to process as many transactions as possible for a given amount of collateral. 
A key performance metric is therefore the \emph{transaction throughput per unit collateral}
where throughput itself is measured either in number of transactions per second or transaction value per second.  


Current PCN designs exhibit poor throughput due to naive design choices in three main areas: (1) \emph{how} to route transactions,(2) \emph{when} to send them and, (3) \emph{deadlocks}.


\NewPara{Challenge \#1: How to route transactions?}
\label{sec:where}
A central question  in  PCNs is what route(s) to use for sending a  transaction  from  sender to destination. 
PCNs like the Lightning and Raiden networks are source-routed.
\footnote{This was done in part for privacy reasons:
transactions in the Lightning network use onion-routing, which is easy to implement with source routing 
\cite{goldschlag1999onion}.}
Most clients by default pick the shortest path from the source to the destination.

However, shortest-path routing degrades throughput in two key ways. 
The first is to cause underutilization of the network. 
To see this, consider the PCN shown in Fig.~\ref{fig:barbell1}. 
Suppose we have two clusters of nodes that seek to transact  with each other at roughly the same rate on average,  
and the clusters are connected by two paths, one consisting of channels $a-b$, and the other channel $c$.
If the nodes in cluster A try to reach cluster B via the shortest path, they would all  take channel $c$, as would the traffic in the opposite direction. This leads to congestion on channel $c$, while channels $a$ and $b$ are
under-utilized.

A second problem is more unique to PCNs. 
Consider a similar topology in Figure \ref{fig:barbell2}, and suppose we fully utilize the network by sending all traffic from cluster A$\to$B on edge $a$ and all traffic from cluster B$\to$A on edge $b$. 
While the rate on both edges is the same,
as funds flow in one direction over a channel, the channel becomes \emph{imbalanced}: all of the funds end up on one side of the channel.
Cluster A can no longer send payments until it receives funds from cluster B on the edge $a$ or 
it deposits new funds into the channel $a$ via an on-chain transaction. 
The same applies to cluster B on edge $b$. 
Since on-chain transactions are expensive and slow, it is desirable to avoid them. 
Routing schemes like shortest-path routing do not account for this problem, thereby leading to reduced throughput (\Sec{sec:eval}).
In contrast, it is important to choose routes that actively prevent channel imbalance. 
For example, in Figure \ref{fig:barbell2}, we could send half of the A$\to$B traffic on edge a, and half on edge $b$, and the same for the B$\to$A traffic.
The challenge is making these decisions in a fully decentralized way.

\NewPara{Challenge \#2: When to send transactions?} 
\label{sec:when}
Another problem is \emph{when} to send transactions. 
Most existing PCNs are circuit-switched: transactions are processed instantaneously and atomically upon arrival \cite{lightning,raiden}.
This causes a number of problems.
If a transaction's value exceeds the available balance on each path from the source to the destination, the transaction fails. 
Since transaction values in the wild tend to be heavy-tailed \cite{kaggledata,cordi2017simulating}, 
either a substantial fraction of real transactions will fail as PCN usage grows, or payment channel operators will need to provision higher 
collateral to satisfy demand.

Even when transactions do not fail outright, sending transactions instantaneously and atomically exacerbates the imbalance problem 
by transferring the full transaction value to one side of the channel.
A natural idea to alleviate these problems is to ``packetize'' transactions: transactions can be split into smaller \tus that can be multiplexed over space (by traversing different paths) and in time (by being sent at different rates).
Versions of this idea have been proposed before; 
atomic multi-path payments (AMP) enable transactions to traverse different paths in the Lightning network \cite{amp}, and the Interledger protocol uses a similar packetization to conduct cross-ledger payments \cite{interledger}. 
However, a key observation is that it is not enough to subdivide transactions into smaller units: to achieve good throughput, 
it is also important to multiplex in \emph{time} as well, by performing congestion control. 
    If there is a large transaction in one direction on a channel,
simply sending it out in smaller units that must all complete together doesn't improve the likelihood of success. Instead, in our design, we allow each transaction-unit to complete independently, and a congestion control algorithm at the sender throttles the rate of these units to match the rate of units in the opposite direction at the bottlenecked payment channel. This effectively allows the tokens at that bottleneck to 
be replenished and reused multiple times as part of the same transaction, achieving a multiplicative
increase in throughput for the same collateral. 

\NewPara{Challenge \#3: Deadlocks.} 
\label{sec:deadlock}
The third challenge in {\PCN}s is the idea that the introduction of 
certain flows can actively harm the throughput achieved by other flows in the network.
To see this, consider the topology and demand rates in Figure \ref{fig:deadlock}. %
Suppose nodes 1 and 2 want to transmit 1-unit transactions to node 3 at rates of 1 and 2 units/second, respectively, and node 3 wants to transact 2 units/sec with node 1.\footnote{For simplicity, we show three nodes, but a node in this example could represent a cluster of many users who wish to transact at the rates shown in aggregate.}
Notice that the specified transaction rates are imbalanced: there  is a net flow of funds out of node 2 and into nodes 1 and 3.
Suppose the payment channels are initially balanced, with 10 units on each side and we only start out with flows between
nodes 1 and 3.
For this demand and topology, the system can sustain 2 units/sec by only having nodes 1 and 3 to send to each other at a rate of 1 unit/second. 

However, once transactions from node 2 are introduced, 
this example achieves zero throughput at steady-state. 
The reason is that node 2  sends transactions to  node 3 faster than  its  funds are being  replenished, which reduces its funds to 0.
Slowing down 2's transactions would only delay this outcome.
Since node 2 needs a positive balance to route transactions between nodes 1 and 3, the transactions between 1 and 3 cannot be processed, despite
the endpoints having sufficient balance. 
The network  finds itself in a \emph{deadlock} that can only be resolved by node 2 replenishing its balance with an on-chain transaction.

\smallskip
\NewPara{Why these problems are difficult to solve.}
The above problems are challenging because their effects  are closely  intertwined.  
For example, because poor routing and rate-control algorithms can cause channel imbalance, which in turn degrades throughput, 
it is difficult to isolate the effects of each. 
Similarly, simply replacing circuit switching with packet-switching gives limited benefits without a  corresponding rate control and routing mechanism. 

From a networking standpoint, PCNs are very different from traditional communication networks: payment channels do not behave like a standard communication link with a certain capacity, say in transactions per second. Instead, the capacity of a channel in a certain direction depends on two factors normally not seen in communication networks: (a) the rate that transactions are received in the reverse direction on that channel, because tokens cannot be sent faster on average in one direction than they arrive in the other, (b) the delay it takes for the destination of a transaction to receive it and send back the secret key unlocking the funds at routers (\S\ref{sec:background}). Tokens that are ``in flight'', i.e. for which a router is waiting for the key, cannot be used to service new transactions. Therefore the network's capacity depends on its delay, and queued up transactions at a depleted link can hold up funds from channels in other parts of the network.
This leads to cascading effects that make congestion control particularly critical.

\eat{We address this problem by formulating a linear program that maximizes transaction throughput while being proportionally-fair. This LP is subject to subject to physical constraints that channel balances should sum up to the overall capacity of each channel, 
and sending rates in each direction over a particular link should be balanced. The resulting \name~protocol that provably solves this LP involves PCN routers 
marking transactions if the observed queuing delay through them is higher than a threshold. Based on the observed marks across
all links on a path, users adjust the transmission rate for that path.}


\section{Packet-Switched PCN}
\label{sec:architecture}

\name~uses a packet-switched architecture that splits transactions into a series of independently routed {\em \tus}. 
Each \tu transfers a small amount of money bounded by a {\em maximum-transaction-unit (MTU)} value. 
Packetizing transactions is inspired by packet switching for the Internet, 
which is more effective than circuit switching \cite{internethistory}.
Note that splitting transactions does not compromise the security of payments; each \tu can be created 
with an independent secret key.  
As receivers receive and acknowledge \tus, senders can selectively reveal secret keys only for acknowledged \tus
(\Sec{sec:background}). 
Senders can also use proposals like Atomic Multi-Path Payments (AMP) \cite{amp} if they desire atomicity of transactions.

In \name, payments transmitted by source {\em end-hosts} are forwarded to their destination end-hosts by {\em routers} within the \PCN.
\name routers queue up \tus at a payment channel whenever the channel lacks the funds to forward them immediately. 
As a router receives funds from the other side of its payment channel, it uses these funds to forward \tus 
waiting in its queue.  
Current PCN implementations~\cite{lightning-implementation} do not queue transactions at routers---a transaction 
fails immediately if it encounters a channel with insufficient balance on its route. 
Thus, currently, even a temporary lack of channel balance can cause many transactions to fail, which \name avoids.

\section{Modeling Routing}
\label{sec:model}

A good routing protocol must satisfy the following objectives: 
\begin{enumerate}[noitemsep,topsep=0pt,parsep=0pt,partopsep=0pt]
\item{\bf Efficiency.}
For a PCN with a fixed amount of escrowed capital, the aggregate transaction throughput achieved must be as high as possible. 
\item {\bf Fairness.}
The throughput allocations to different users must be fair. 
Specifically, the system should not starve transactions of some users if there is capacity.
\end{enumerate}

Low latency, a common goal in communication networks, is desirable but not a first order concern, as long as transaction latency on the PCN is significantly less than an on-chain transaction (which can take minutes to hours today). However, as mentioned previously (\S\ref{sec:motivation}), very high latency could hurt the throughput of a PCN, and must therefore be avoided. We assume that the underlying communication network is not a bottleneck and \PCN users can communicate payment attempts, success and failures with one another easily since these messages do not require much bandwidth.

To formalize the routing problem, 
we consider a fluid model of the system in which payments are modeled as continuous ``fluid flows'' between users. 
This  allows us to cast routing as an optimization problem and derive  decentralized algorithms from it, analogous to the classical Network Utility Maximization (NUM) framework for data networks~\cite{palomar2006tutorial}.
More specifically, for the fluid model we consider a PCN modeled as a graph $G(V,E)$ in which $V$ denotes the set of nodes (i.e., end-hosts or routers), and $E$ denotes the set of payment channels between them. 
For a path $p$, let $x_p$ denote the (fluid) rate at which payments are sent along $p$ from a source to a destination.
The fluid rate captures the long-term average rate at which payments are made on the path. 

For maximizing throughput efficiency, routing has to be done such that the total payment flow through each channel is as high as possible. 
However, routers have limited capital on their payment channels, which restricts the maximum rate at which funds can be routed (\Fig{barbell1}). 
In particular, when transaction units are sent at a rate $x_{u,v}$ across a payment channel between $u$ and $v$ with $c_{u,v}$ funds in total and it takes $\Delta$ time units on average to receive the secret key from a destination once a payment is forwarded, then $x_{u,v} \Delta$ credits are locked (i.e., unavailable for use) at any point in time in the channel. 
This implies that the average rate of transactions (across both directions) on a payment channel cannot exceed
$c_{u,v} / \Delta$. 
This leads to {\em capacity constraints} on channels. 


Sustaining a flow in one direction through a payment channel requires funds to be regularly replenished from the other direction. 
This requirement is a key difference between PCNs and traditional data networks. 
In {\PCN}s if the long-term rates $x_{u,v}$ and $x_{v, u}$ are mismatched on a channel $(u,v)$, say $x_{u, v} > x_{v, u}$, then over time all the funds $c_{u, v}$ will accumulate at $v$ deeming the channel unusable in the direction $u$ to $v$ (\Fig{barbell2}). 
This leads to {\em balance constraints} which stipulate that the total rate at which transaction units are sent in one direction along a payment channel matches the total rate in the reverse direction.  

Lastly, for enforcing fairness across flows we assume sources have an intrinsic {\em utility} for making payments, which they seek to maximize. 
A common model for utility at a source is the logarithm of the total rate at which payments are sent from the source~\cite{kelly1998rate, kelly2005stability, eryilmaz2006joint}.
A logarithmic utility ensures that the rate allocations are proportionally fair~\cite{kelly1998rate}---no individual sender's payments can be completely throttled.   
Maximizing the overall utility across all source-destination pairs subject to the capacity and balance constraints discussed above, can then be computed as
\begin{align}
\mathrm{maximize}  \qquad \sum_{i,j \in V} \log & \Big(\sum_{p\in\mathcal{P}_{i,j}} x_p\Big) \label{eq:LP} \\
\mathrm{s.t.} ~~~ \quad \quad \qquad \sum_{p\in\mathcal{P}_{i,j}} x_p &\leq d_{i, j} \quad \forall i, j \in V 
\label{eq:LP demand}  \\
x_{u, v}  + x_{v, u} &\leq \frac{c_{u, v}}{\Delta} \quad \forall (u,v) \in E \label{eq:LP capacity} \\
x_{u, v} &= x_{v, u} \quad \forall (u,v) \in E \label{eq:LP balance} \\
x_p &\geq 0 \quad \forall p \in \mathcal{P},   \label{eq:LP end}
\end{align}
where for a source $i$ and destination $j$, $\mathcal{P}_{i,j}$ is the set of all paths from $i$ to $j$, $d_{i,j}$ is the demand from $i$ to $j$, $x_{u,v}$ is the total flow going from $u$ to $v$ for a channel $(u,v)$, $c_{u,v}$ is the total amount of funds escrowed into $(u,v)$, $\Delta$ is the average round-trip time of the network taken for a payment to be completed, and $\mathcal{P}$ is the set of all paths. 
Equation~\eqref{eq:LP demand} specifies {\em demand constraints} which ensures that the total flow for each sender-receiver pair across all of their paths, is no more than their demand. 




\begin{figure}
    \centering
    \begin{tabular}{c c c}
        \begin{subfigure}{0.2\columnwidth}
            \centering
            \includegraphics[width=\columnwidth]{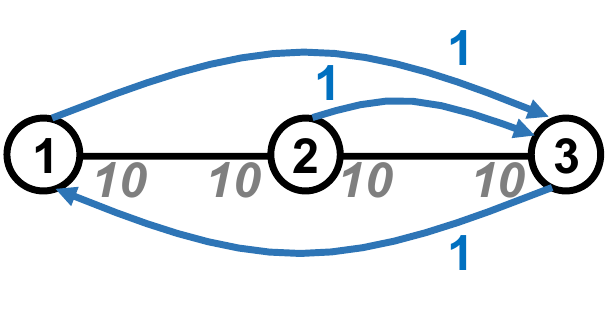}
            \subcaption{Payment graph}
\label{fig:payment_graph_1}
        \end{subfigure} &
        \begin{subfigure}{0.2\columnwidth}
            \centering
            \includegraphics[width=\columnwidth]{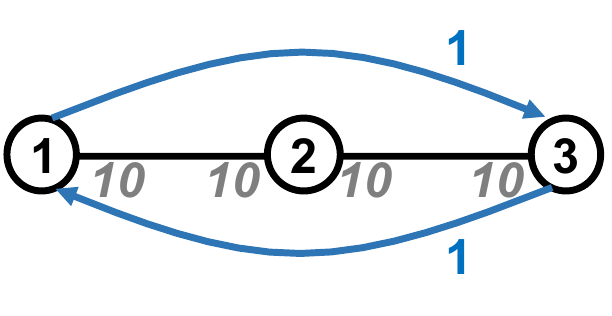}
            \subcaption{Circulation}
\label{fig:payment_graph_2}
        \end{subfigure} &
        \begin{subfigure}{0.2\columnwidth}
            \centering
            \includegraphics[width=\columnwidth]{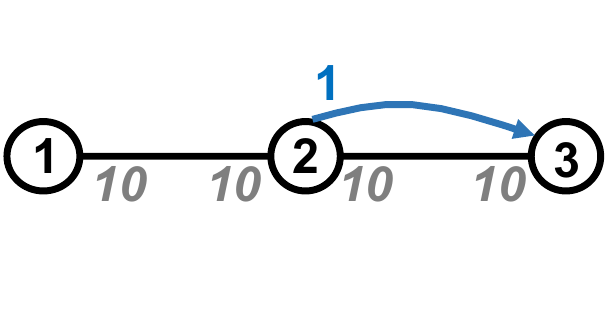}
            \subcaption{DAG}
\label{fig:payment_graph_3}
        \end{subfigure}         
    \end{tabular}
    \caption{\small Payment graph (denoted by blue lines) for a 3 node network (left). It decomposes into a maximum circulation and DAG components as shown in (b) and (c).}
\label{fig:payment_graph}
\vspace*{-3mm}
\end{figure}

\subsection{Implications for Throughput} 
\label{sec:throughput bounds}

A consequence of the balance constraints is that certain traffic demands are more efficient to route than certain others. 
In particular, demands that have a \emph{circulation} structure (total outgoing demand matches total incoming demand at a router) can be routed efficiently. 
The cyclic structure of such demands enables routing along paths such that the rates are naturally balanced in channels. 
However, for demands without a circulation structure, \ie if the demand graph is a directed acyclic graph (DAG), balanced routing is impossible to achieve in the absence of periodic replenishment of channel credits, regardless of how large the channel capacities are. 
 
For instance, \Fig{payment_graph_1} shows the traffic demand graph for a PCN with nodes $\{1, 2, 3\}$ and payment channels between nodes $1-2$ and $2-3$. 
The weight on each blue edge denotes the demand in \tus per second 
between a pair of users. The underlying black lines denote the topology and channel sizes. 
\Fig{payment_graph_2} shows the circulation component of the demand in \Fig{payment_graph_1}. 
The entire demand contained in this circulation can be routed successfully as long as the network
has sufficient capacity. 
In this case, if the confirmation latency for \tus between 1 and 3 is less than 10s, then the circulation demand can be satisfied indefinitely.
The remaining component of the demand graph, which represents the DAG, is shown in \Fig{payment_graph_3}. 
This portion cannot be routed indefinitely since it shifts all tokens onto node 3 after which the $2-3$ channel
becomes unusable.

\App{apx: circulation_proof} formalizes the notion of circulation and shows that the maximum throughput achievable by any balanced routing scheme is at most the total demand contained within the circulation. 

\subsection{A Primal-Dual Decomposition Based Approach}
We now describe a decentralized algorithm based on standard primal-dual decomposition techniques
used in utility-maximization-based rate control and routing literature (e.g.~\cite{kelly2005stability}).
\Sec{sec:pd-router} and \Sec{sec:pd-sender} discuss the protocol at router nodes and end-hosts in order
to facilitate this approach. A detailed derivation of the algorithm using in the fluid model 
that also considers the cost of on-chain rebalancing 
is discussed in \App{apx: primal dual derivation}. However, \Sec{sec:pd-challenges} outlines
the difficulties with this approach that lead us to the design of the practical protocol 
discussed in \Sec{sec:protocol}.

To arrive at this algorithm, we consider the optimization problem described in \Sec{sec:model} for a
generic utility function $U(\sum_{p\in\mathcal{P}_{i,j}} x_p)$. 
The structure of the Lagrangian of the LP allows us to naturally decompose the overall optimization problem into separate subproblems for each sender-receiver pair. 

A solution to this LP can be computed via a decentralized algorithm in which each sender maintains rates
at which payments are sent on each of its paths.
Each payment channel has a {\em price} in each direction. 
Routers locally update these prices depending on both congestion and imbalance across the payment channels, while end-hosts adjust their rates by monitoring the total price on the different paths.
The primal variables of the LP represent the rate of payments on each path, and the dual variables represent the channel prices. While this approach has been used before ~\cite{kelly2005stability}, a key difference from prior work is the presence of price variables for link balance constraints in addition to the price variables for capacity constraints. 
This ensures that the price of a channel having a skewed balance is different in each direction, and steers the flow rates to counter the skew. 

\subsubsection{Router Design} 
\label{sec:pd-router}
~~ \\
Routers in each payment channel under this algorithm maintain price variables, which are updated periodically based on the current arrival rate of transaction units in the channel, available channel balance, and the number of transaction units queued up at the routers. 
The price variables at the routers determine the path prices, which in turn affect the rates at which end-hosts transmit transaction units (we discuss more in \Sec{sec:pd-sender}). 

In a payment channel $(u,v) \in E$, routers $u$ and $v$ hold estimates for three types of price variables: $\lambda_{u,v}, \mu_{u,v}$ and $\mu_{v,u}$.
These are dual variables corresponding to the capacity and imbalance constraints in Equation~\eqref{eq:LP balance} and~\eqref{eq:LP capacity} respectively.
The \emph{capacity price} $\lambda_{u,v}$ signals congestion in the channel if the total rate at which transactions arrive exceeds its  capacity; the \emph{congestion prices}   $\mu_{u,v}$ and $\mu_{v,u}$ are used to signal imbalance in the channel in the two directions. 
These variables are updated periodically to ensure that the capacity and imbalance conditions are not violated in the channel.
Prices are updated every $\tau$ seconds according to the rules described next. 

\NewPara{Imbalance Price.} For a channel $(u,v)$, let $n_u, n_v$ denote the total amount of transactions that have arrived at $u$ and $v$ respectively, in the $\tau$ seconds since the last price update. 
The price variable for imbalance $\mu_{(u,v)}$ is updated as 
\begin{align}
\mu_{u,v}(t+1) = \left[ \mu_{u,v}(t) + \kappa \left(n_u(t) - n_v(t) \right) \right]_+, \label{eq:mu update sys}
\end{align} 
where $\kappa$ is a positive step-size parameter for controlling the rate at which the price varies.
\footnote{The price update for $\mu_{v,u}$ is analogous to Eq.~\eqref{eq:mu update sys}, but with $u$ and $v$ interchanged. This equation can
be modified to include on-chain rebalancing amounts on both ends.} Intuitively, if more funds arrive in the $u$-$v$ direction compared to the $v$-$u$ direction (i.e., $n_u(t) > n_v(t)$), the price $\mu_{u,v}$ increases while the price $\mu_{v,u}$ decreases. 
The higher price in the $u$-$v$ direction signals end-hosts that are routing along $(u,v)$ to throttle their rates, and signal end-hosts routing along $(v,u)$ to increase their rates.


\NewPara{Capacity Price.} The price variable for capacity $\lambda_{u,v}$ is also updated every $\tau$ seconds as follows:
\begin{align}
\lambda_{u,v}(t+1) = \left[ \lambda_{u,v}(t) + \eta \left( m_u(t) + m_v(t) - c_{u,v} \right) \right]_+. \label{eq:lambda update sys}
\end{align}
For the current rates of transaction arrival at $u$ and $v$, $m_u(t)$ and $m_v(t)$ are estimates of the amount of funds required to sustain those rates at $u$ and $v$ respectively.  
Since $c_{u,v}$ is the total amount of funds available in the channel, any excess in required amount of funds compared to $c_{u,v}$ would cause $\lambda_{u,v}$ to rise and vice-versa. 
An increase in $\lambda_{u,v}$ signals end-hosts routing via $u,v$, on either direction, to reduce their rates.  
We estimate the demands $m_u(t)$ and $m_v(t)$ for tokens by measuring the arrival and service rates of transactions, and the current amount of locked funds in the channel as described in \App{apx: demand estimation}.
$\eta$ is a positive step-size parameter.  

\subsubsection{End-host Design} 
\label{sec:pd-sender}
~~ \\
\TheSystem-hosts run a multi-path transport protocol with
pre-determined paths which controls the rates at which payments are transferred, based on observations of the channel prices or router feedback. End-hosts here use probe messages to evaluate the channel prices on each path. The total price of a path $p$ is given by
\begin{align}
z_p = \sum_{(u,v): (u,v)\in p} (2\lambda_{u,v} + \mu_{u,v} - \mu_{v,u}), \label{eq: path price}
\end{align}
which captures the aggregate amount of imbalance and excess demand, as signaled by the corresponding price variables, in the path.
We refer to \App{apx: primal dual derivation} for a mathematical intuition behind Equation~\eqref{eq: path price}. 
Probes are sent periodically every $\tau$ seconds (i.e., the same frequency at which channel prices are updated \S\ref{sec:router design}) on each path.
A probe sent out on path $p$ sums the price $2\lambda_{u,v} + \mu_{u,v} - \mu_{v,u}$ of each channel $(u,v)$ it visits, until it reaches the destination host on $p$.
The destination host then transmits the probe back to the sender along the reverse path. 
The rate to send $x_p$ on each path $p$ is updated using the path price $z_p$ from the most recently received probe as   
\begin{align}
    x_p(t+1) = x_p(t) + \alpha(U'(x) - z_p(t)), \label{eq:rate update sys}
\end{align}
where $\alpha$ is a positive step-size parameter. 
Thus the rate to send on a path decreases if the path price is high---indicating a large amount of imbalance or capacity deficit in the path---and increases otherwise. 

\subsubsection{Challenges}
\label{sec:pd-challenges}
There are a number of challenges in making this algorithm work in practice. 
Firstly, iterative algorithms for adjusting path prices and sending rates suffer from slow convergence. An algorithm that is slow to converge may not be able to adapt the routing to changes in the transaction arrival pattern. If the transaction arrival patterns change frequently, this may result in a perpetually suboptimal routing.
Secondly, in order to compute the
imbalance prices, the two routers in a payment channel need to exchange information about their arrival patterns
and their respective queue states to calculate $n_u$ and $m_u$ in 
Eq.\ref{eq:mu update sys}--\ref{eq:lambda update sys}. This implies that routers cannot deploy this
in isolation.
Further, we found that the scheme was extremely sensitive to the many parameters involved in the algorithm, making it hard to tune for a variety of topologies and capacity distributions. Lastly, a pure pacing based approach 
could cause bursts in \tus sent
that lead to large queue buildups much before
the prices react appropriately. To account for this, the algorithm needs to be augmented with windows \cite{vanjacobson}  
and active queue control to work in practice. 
Due to these difficulties, we propose a more practical and simpler protocol described in \Sec{sec:protocol}.

\section{Design}
\label{sec:protocol}

\subsection{Intuition}
\label{sec:intuition}

\begin{figure}
    \begin{subfigure}{\columnwidth}
	\includegraphics[width=0.68\columnwidth]{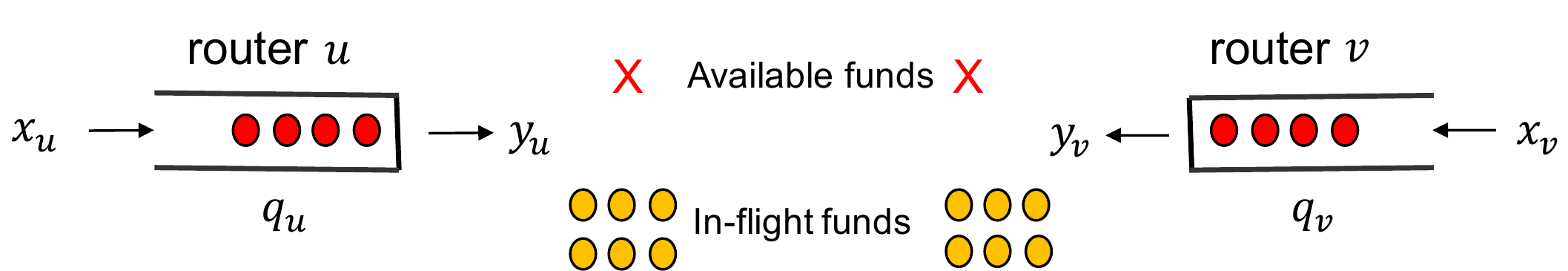}
	\centering
        \caption{\small A capacity limited payment channel.}
        \label{fig:capacity_cartoon}

    	\includegraphics[width=0.68\columnwidth]{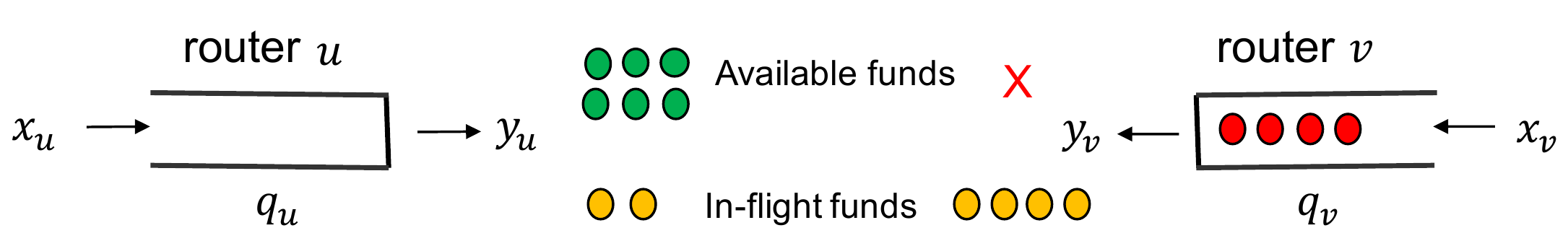}
        \caption{\small An imbalance limited payment channel. } 	
        \label{fig:imbalance_cartoon}
    \end{subfigure}
    \label{fig:cartoon}
    \caption{\small Example of queue growth in a payment channel between routers $u$ and $v$, under different scenarios of transaction arrival rates at $u$ and $v$. (a) If the rate of arrival at $v$, $x_v$, and the rate of arrival at $u$, $x_u$, are such that their sum exceeds the channel capacity, neither router has available funds and queues build up at both $u$ and $v$. 
    (b) If the arrival rates are imbalanced, \eg if $x_v > x_u$, then $u$ has excess funds while $v$ has none, causing queue build-up at $v$. }
\end{figure}
%

\TheSystem routers queue up transactions at a payment channel whenever the channel lacks funds to forward them immediately (\Sec{sec:model}). 
Thus, queue buildup is a sign that either \tus are arriving faster (in both directions) than the channel can process (\Fig{capacity_cartoon}) or 
that one end of the payment channel lacks sufficient funds(\Fig{imbalance_cartoon}). It indicates that
the capacity constraint (\Eqn{LP capacity}) or the balance constraint (\Eqn{LP balance}) is being violated and
the sender should adjust its sending rate.



Therefore, if senders use a congestion control protocol that controls queues, they could detect both capacity and imbalance violations and react to them. For example, in \Fig{capacity_cartoon}, the protocol would throttle both $x_u$ and $x_v$.
In \Fig{imbalance_cartoon}, it would decrease $x_v$ to match the rate at which queue $q_v$ drains, which is precisely $x_u$, the rate at which new funds become available at router $v$.

This illustrates that a congestion controller that satisfies two basic properties can achieve both efficiency and balanced rates:  
\begin{enumerate}[noitemsep,topsep=0pt,parsep=0pt,partopsep=0pt]
\item {\em Keeping queues non-empty,} which ensures that any available capacity is being utilized, \ie there are no unused tokens at any router. 

\item {\em Keeping queues stable (bounded),} which ensures that (a) the flow rates do not exceed a channel's capacity, (b) the flow rates are balanced. If either condition is violated, then at least one of the channel's queues would grow. 
\end{enumerate}
Congestion control algorithms that satisfy these properties abound (\eg Reno~\cite{reno}, Cubic~\cite{ha2008cubic}, DCTCP~\cite{dctcp}, Vegas~\cite{brakmo1994tcp}, etc.) and could be adapted for PCNs.

In PCNs, it is desirable to transmit \tus along multiple paths to better utilize available capacity. Consequently, \name's design is inspired by multi-path transport protocols like MPTCP~\cite{MPTCP}. These protocols couple rate control decisions for multiple paths  to achieve both high throughput and fairness among competing flows~\cite{wischik2008resource}. We describe an MPTCP-like protocol for PCNs in \S\ref{sec:router design}--\ref{sec:transport layer endhosts}. 
In \S\ref{sec:dctcpproof} we show that the rates found by \name's protocol for parallel network topologies, match the solution to the optimization problem in \S\ref{sec:model}.

\subsection{\name Router Design}
\label{sec:router design}
\begin{figure}[t]
\begin{center}
\includegraphics[width=3.3in]{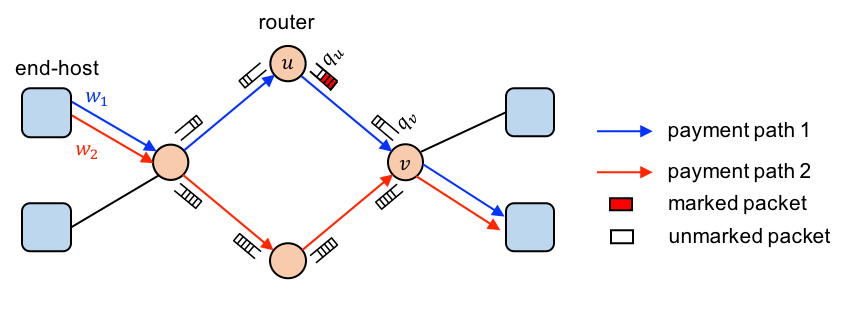}
\caption{\small Routers queue up \tus and schedule them based on priorities when funds become available. and transaction priorities. If the delay through the queue for a packet exceeds a threshold, 
    they mark the packet.
End-hosts maintain and adjust windows for each path to a receiver based on the marks they observe.
}
\label{fig:systemoverview}
\end{center}
\vspace*{-6mm}
\end{figure}

\Fig{systemoverview} shows a schematic diagram of the various components in the \TheSystem PCN.
\name routers monitor the time that each packet spends in their queue and mark the packet if 
the time spent exceeds a pre-determined threshold $T$.
If the \tu is already
marked, routers leave the field unchanged and merely forward the \tu.  Routers forward acknowledgments from the receiving end-host back to the sender 
which interprets the marked bit in the ack accordingly. \name routers schedule \tus from their queues according to a scheduling policy, like Smallest-Payment-First or Last-In-First-Out (LIFO). Our evaluations (\Sec{sec:design choices}) shows that LIFO provides the highest transaction success rate. The idea behind LIFO is to prioritize transaction units from new payments, which are likely to complete within their deadline.


\subsection{\name Transport Layer at End-Hosts}
\label{sec:transport layer endhosts}
\name senders send and receive payments on a PCN by interfacing with their transport layer. This layer is configured to support both atomic and non-atomic payments
depending on user preferences.
Non-atomic payments utilize \name's packet-switching which breaks up large payments into \tus that are delivered to the receiver independently.  
In this case, senders are notified of how much of the payment was completed allowing them to cancel the rest or retry it on the 
blockchain.  
While this approach crucially allows token reuse at bottleneck payment channels for the same transaction (\S\ref{sec:motivation}), senders also have the option of requesting atomic payments (likely for a higher fee). 
Our results (\Sec{sec:eval}) show that even with packetization, more than 95\% payments complete in full

The transport layer also involves a multi-path protocol which controls the rates at which payments are transferred, based on congestion in the network. 
For each destination host, a sender chooses a set of $k$ paths to route \tus along.
The route for a \tu is decided at the sender before transmitting the unit.
It is written into the \tu using onion encryption, to hide the full route from intermediate routers \cite{goldschlag1999onion,onionlightning}. 
In \Sec{sec:design choices}, we evaluate the impact of different path choices on \TheSystem's performance and propose using edge-disjoint widest paths \cite{networkflows} between each sender and receiver in \TheSystem.  

To control the rate at which payments are sent on a path, end-hosts maintain a window size $w_p$ for 
every candidate path to a destination. This window size denotes the maximum number of \tus that can be outstanding 
on path $p$ at any point in time. End-hosts track the \tus that have been sent out on each path but have not yet been acked or 
canceled. A new \tu is transmitted on a path $p$ only if the total amount pending does not exceed $w_p$. 

End-hosts adjust $w_p$ based on router feedback on congestion and imbalance.
In particular, on a path $p$ between source $i$ and receiver $j$ the window changes as 
\begin{align}
    w_p &\leftarrow w_p - \beta, &\text{on every marked packet and,} \label{eq:window dec} \\
    w_p &\leftarrow w_p +\frac{\alpha}{\displaystyle \sum_{p': p' \in \mathcal{P}_{i,j}}w_{p'}},  &\text{on every unmarked packet}\label{eq:window inc}.
\end{align}
Here, $\alpha$ and $\beta$ are both positive constants that denote the aggressiveness with which the
window size is increased and decreased respectively. Eq.~\eqref{eq:window dec}--\eqref{eq:window inc} are similar to MPTCP, but with a multiplicative decrease factor that depends on the fraction of packets marked on a path (similar to DCTCP~\cite{dctcp}). 

We expect the application to specify a deadline for every transaction. If the transport layer
fails to complete the payment within the deadline, the sender cancels the payment,
clearing all of its state from the PCN. In particular, it sends a cancellation message to remove any \tus queued at routers on each path to the receiver. Notice that \tus that arrive at the receiver in the meantime cannot be unlocked because we assume the sender holds the secret key (\S\ref{sec:background}). Senders can then choose to retry the failed portion of the transaction again on the PCN or on the blockchain; such retries would be treated as new transactions.
Canceled packets are considered marked and \name
decreases its window in response to them.


\subsection{Optimality of \name}
\label{sec:dctcpproof}

Under a fluid approximation model for \name's dynamics, we can show that the rates computed by \name are an optimal solution to the routing problem in Equations~\eqref{eq:LP}--\eqref{eq:LP end} for parallel networks (such as \Fig{mptcp_proof} in \App{apx:fluidmodel}). 
In the fluid model, we let $x_p(t)$ denote the rate of flow on a path $p$ at time $t$; for a channel $(u,v)$, $f_{u,v}(t)$ denotes the fraction of packets that are marked at router $u$ as a result of excessive queuing.
The dynamics of the flow rates $x_p(t)$ and marking fractions $f_{u,v}(t)$ can be specified using   differential equations to approximate the window update dynamics in Equations~\eqref{eq:window dec} and~\eqref{eq:window inc}.
We elaborate more on this fluid model, including specifying how the queue sizes and marking fractions evolve, in \App{apx:fluidmodel}. 

Now, consider the routing optimization problem (Equations~\eqref{eq:LP}--\eqref{eq:LP end}) written in the context of a parallel network. 
If \name is used on this network, we can show that there is a mapping from the rates $\{x_p\}$ and marking fractions $\{f_{u,v}\}$ values after convergence, to the primal and dual variables of the optimization problem, such that the Karush-Kuhn-Tucker (KKT) conditions for the optimization problem are satisfied.  
This proves that the set of rates found by \name is an optimal solution to the optimization problem~\cite{boyd2004convex}. 
The complete and formal mathematical proof showing the above is presented in \App{apx:fluidmodel}.

\section{Evaluation} 
\label{sec:eval}


We develop an event-based simulator for {\PCN}s, 
and use it to extensively evaluate \name across a wide range of scenarios. 
We describe our simulation setup (\Sec{sec:setup}), validate it via a prototype implementation (\Sec{sec:impl}), and present detailed results for circulation demands (\Sec{sec:circulation}).   
We then show the effect of adding DAG components to circulations (\Sec{sec:dag}),
 and study {\name}'s design choices (\Sec{sec:design choices}).


\subsection{Experimental Setup}

\label{sec:simulator}
\noindent\textbf{Simulator.} We extend the OMNET++ simulator (v5.4.1)~\cite{omnet} to 
model a \PCN.
Our simulator  
accurately models the network-wide effects of transaction processing, 
by explicitly passing messages between \PCN nodes (endhosts and routers).\footnote{https://github.com/spider-pcn/spider-omnet} 
Each endhost (i) generates transactions destined for other endhosts as per the specified workload, 
and (ii) determines when to send a transaction and along which path, as per the specified routing scheme.
All endhosts maintain a view of the entire \PCN topology, to compute suitable source-routes.  
The endhosts can't view channel balances, but they do know each channel's size or total number of tokens (\eu).
Endhosts also split generated transactions into MTU-sized segments (or \tus) 
before routing, if required by the routing scheme (e.g. by \name). 
Each generated transaction has a \emph{timeout} value and is marked as a failure if it fails to reach its destination by then.
Upon receiving a transaction, an endhost generates an acknowledgment 
that is source-routed along its reverse path. 

A router forwards incoming transactions and acknowledgments along the payment channels
specified in their route, while correspondingly decrementing or incrementing the channel balances.
Funds consumed by a transaction in a channel are \emph{inflight} and unavailable until
its acknowledgment is received.
A transaction is forwarded on a payment channel only if the channel has sufficient balance; otherwise the transaction is stored in a \emph{per-channel queue}
that is serviced in a last in first out (LIFO) order \Sec{sec:design choices}. 
If the queue is full, an incoming transaction is dropped, 
and a failure message is sent to the sender. 

\label{sec:schemes}
\noindent{\bf Routing Schemes.} We implement and evaluate  five different routing schemes in our simulator.

\noindent\emph{(1) \name:} 
Every \name sender maintains a set of up to $k$ edge-disjoint widest paths to each 
destination and a window size per path. 
The sender splits transactions into \tus and sends a \tu on a path
if the path's window is larger than amount inflight on the path.
If a \tu cannot be sent, it is placed in a per-destination queue at the sender
that is served in LIFO order. \name routers mark \tus experiencing queuing delays
higher than a pre-determined threshold. \name receivers echo the mark back to 
senders who adjust the window size according to the equations in \Sec{sec:transport layer endhosts}. 

\noindent\emph{(2) Waterfilling:} 
\WF uses balance information explicitly in contrast to \name's 1-bit feedback.
As with \name, a sender splits transactions into \tus and picks up to $k$ edge-disjoint 
widest paths per destination. 
It maintains one outstanding probe per 
path that computes the bottleneck (minimum) channel balance along it. 
When a path's probe is received, the sender computes the available balance based on its bottleneck and the in-flight \tus.
A \tu  is sent along the path 
with the highest available balance. 
If the available balance for all of the $k$ paths is zero (or less), the \tu
is queued and retried after the next probe. 

\noindent\emph{(3) Shortest Path:} 
This baseline sends transactions  along the shortest path to the destination without transaction splitting.

\noindent\emph{(4) Landmark Routing:} Landmark routing, as used in prior PCN routing schemes~\cite{prihodko2016flare, malavolta2016silentwhispers, speedymurmurs},
chooses $k$ well-connected \emph{landmark} nodes in the topology.
For every transaction, the sender computes its shortest path to each landmark and concatenates it
with the shortest path from that landmark to the destination to obtain $k$ distinct paths. 
Then, the sender probes each path to obtain its bottleneck balance, and partitions the transaction such that each path can support its share of the total transaction. 
If such a partition does not exist or if any of the partitions fail, the transaction fails. 

\noindent\emph{(5) LND:} 
The \PCN scheme  currently deployed in the Lightning Network 
Daemon (LND)~\cite{lightning-implementation} attempts first send a transaction along the shortest path to its destination.
If the transaction fails due to insufficient balance at a channel, 
the sender removes that channel from its local view, 
recomputes the shortest path, and retries the transaction on the new path
until the destination becomes unreachable or the transaction times out.
A channel is added back to the local view 5 seconds after its removal. 

\noindent\emph{(6) Celer:} \App{app:celer} compares \name to Celer's cRoute as proposed in a
white-paper \cite{celer}. Celer is a back-pressure routing algorithm that routes transactions based on queue and imbalance gradients. 
Due to computation overheads associated with Celer's large queues, we evaluate it on a smaller topology.

\eat{
\NewPara{Primal-Dual Approach}: We evaluate the primal-dual based approach described in \Sec{sec:primaldual}
 to compare its ability to achieve fairness relative to \name. We introduce
new messages between neighbors to exchange the number of transactions that arrive in each direction
on a payment channel. These messages are exchanged every 1.5 seconds and are used to update congestion
and imbalance prices (Eqns.~\eqref{eq:lambda update sys}--\eqref{eq:mu update sys}).
These prices are then used to compute the path price that is relayed
to end-hosts in response to probe messages every 1.5 seconds. Senders use the path prices to 
update their sending rates as per \Eqn{rate update sys}. These rates are projected onto the estimated demands
at every sender.
The sending rate along each path is multiplied with
its two-way propagation delay to compute the sending window for the path. 
The senders pace the \tus of a transaction based on the computed sending rates using independent timers on each 
of the $k$ edge-disjoint widest paths, and ensure that
the in-flight transaction amount along a path does not exceed its window.
}

\label{sec:setup} 

\NewPara{Workload.} 
We generate two forms of payment graphs to specify the rate at which a sender transacts with every other receiver: (i) pure circulations, with a fixed total sending rate $x$ per sender. The traffic demand matrix for this is generated by adding $x$ random permutation matrices; (ii) circulations with a DAG component, having a total rate $y$. 
This type of demand is generated by sampling $y$ different sender-receiver pairs where the senders and receivers are chosen from two separate exponential distributions (so that some nodes are more likely than others to be picked as a sender or receiver). 
The scale $\beta$ of the distribution is set proportional to the desired percentage of DAG component in the total
traffic matrix: the greater the fraction of DAG component desired, the more skewed the distribution  becomes.
\begin{figure}
    \centering
    \begin{subfigure}[t]{0.35\columnwidth}
    	\includegraphics[width=0.8\columnwidth]{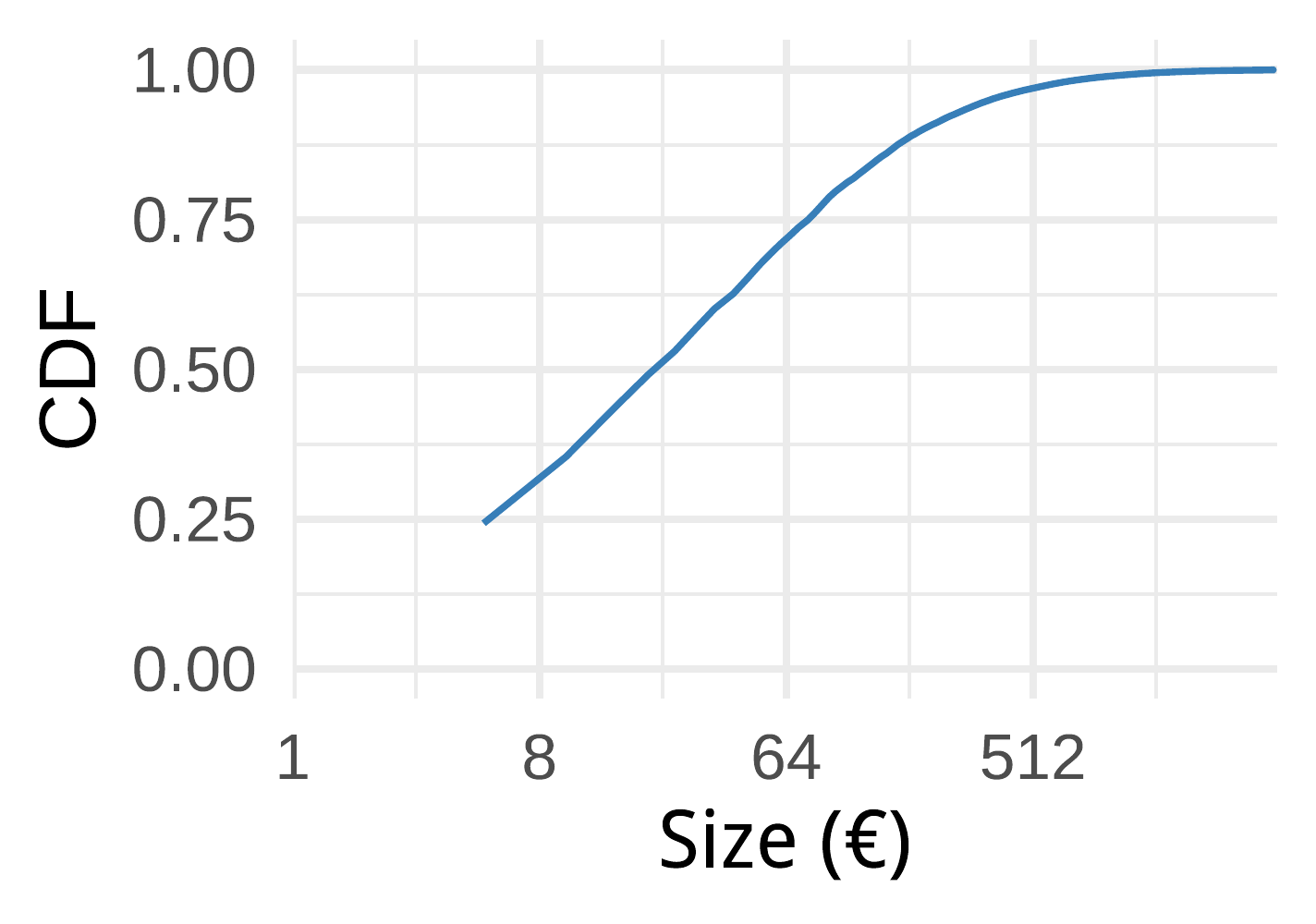}
        \caption{\small Transaction Size Distribution}
   	\label{fig:kaggle}
    \end{subfigure}
    \begin{subfigure}[t]{0.35\columnwidth}
    	\includegraphics[width=0.8\columnwidth]{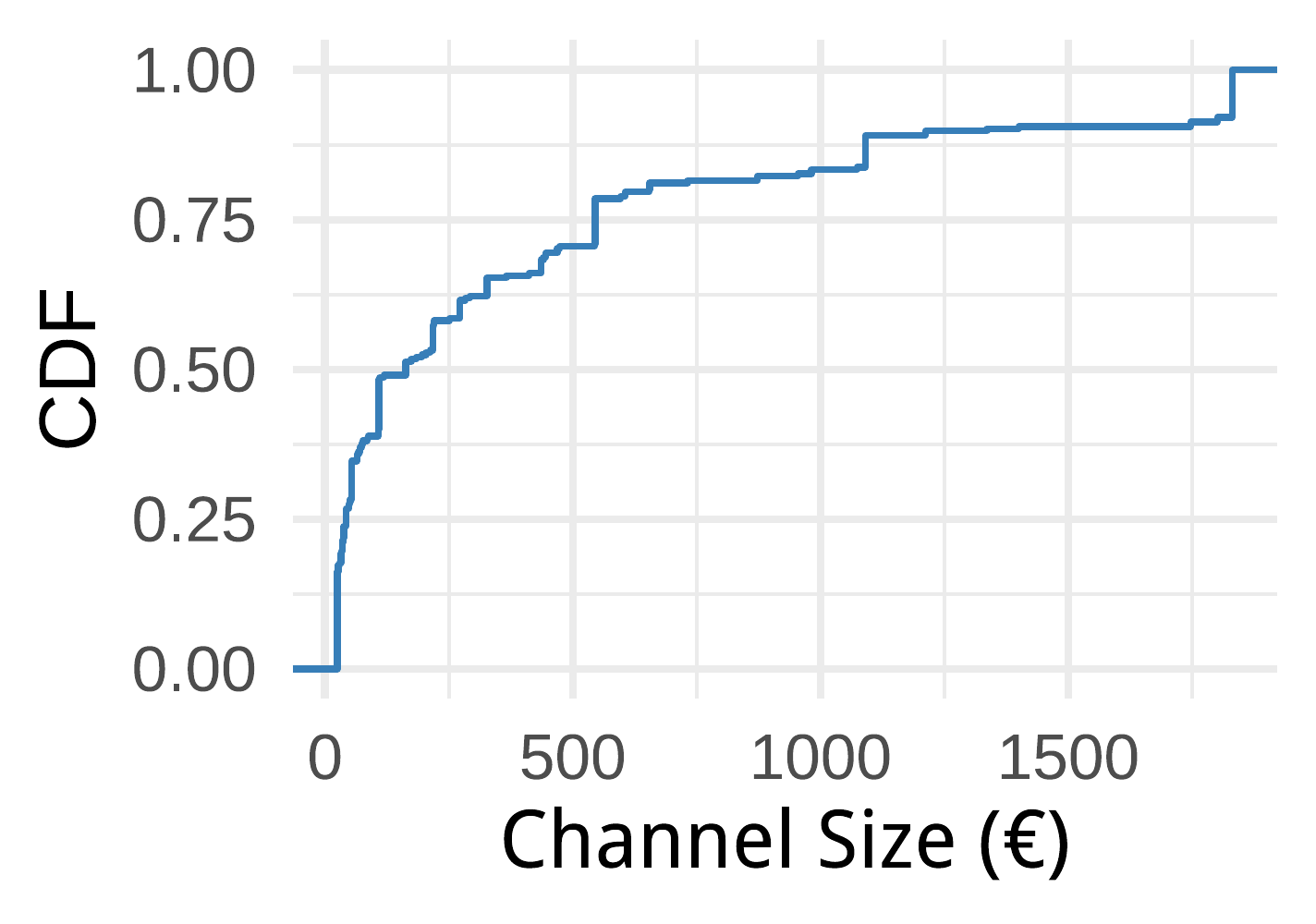}
        \caption{\small LN Channel Size Distribution}
   	\label{fig:lnd:topo:capacity}
    \end{subfigure}~
		\caption{\small Transaction dataset and channel size distribution used for
                real-world evaluations.}
    \label{fig:dataset}
\end{figure}
We translate the rates specified in the payment graphs to discrete transactions with Poisson inter-arrival times.
The transaction size distribution is drawn from credit card transaction data \cite{kaggledata}, 
and has a mean of 88\eu and median 25\eu with
the largest transaction being 3930\eu. The distribution of transaction sizes is shown in \Fig{kaggle}. 
We keep the sending rates constant at an average of 30 tx/sec per sender that is shared
among 10 destinations throughout all of our experiments. Note that a sender represents a router in these experiments, sending transactions to other routers on behalf of many users.


\NewPara{Topology.}
We set up an LND node~\cite{lightning-implementation} to retrieve the the Lightning Network topology on July 15, 2019. We snowball sample~\cite{snowball-sampling} the full topology (which has over 
5000 nodes and 34000 edges), resulting in a \PCN with 
106 nodes and 265 payment channels. 
For compatibility with our transaction dataset, we convert LND payment channel sizes from Satoshis to \eu, and cap the minimum channel size to the median transaction size of 25\eu.
The distribution of channel sizes for this topology
has a mean and median size of 421\eu and 163\eu respectively (\Fig{lnd:topo:capacity}).
This distribution is highly skewed, resulting in a mean that is much larger than the median or the smallest
payment channels.
We refer to this distribution as the Lightning Channel Size Distribution (LCSD).
We draw channel propagation delays based on ping times from our LND node to all reachable nodes
in the Lightning Network, resulting in transaction minimum RTTs of about a second.


We additionally simulate two synthetic topologies: a Watts-Strogatz small world topology \cite{smallworld} with 50 nodes and 200 edges, and a scale-free Barabasi-Albert graph \cite{scalefree} with 50 nodes and 336 edges. We set the per-hop delay to 30ms in both cases, resulting in an end-to-end minimum RTT of 200-300ms. 

For payment channel sizes, we use real capacities in the Lightning topology and sample capacities from LCSD for synthetic topologies.
We vary the mean channel size across different experiments by proportionally scaling up the size of each payment channel. 
All payment channels are perfectly balanced at the start of the experiment. 



\NewPara{Parameters.}
We set the MTU as 1\eu. 
Every transaction has a timeout of $5$ seconds. 
Schemes with router queues enabled have a per-channel queue size of 12000\eu. 
The number of path choices is set to $k=4$ for schemes that use multiple paths.
We vary both the number of paths and the nature of paths in \Sec{sec:design choices}. 
For \name, we set $\alpha$ (window increase factor) to $10$, $\beta$ (multiplicative decrease factor) to $0.1$, and the marking threshold for the queue delay to $300$ms. For the experiments in \Sec{sec:dag}, we set 
this threshold to $75$ms so as to respond to queue buildup faster than per-RTT.

\eat{
    For the primal-dual approach from \Sec{sec:primaldual}, 
the frequency of price queries and updates ($\tau$) is set to 1.5s. 
The step size  $\alpha$ for the rate update at senders is set to $0.2$. 
The step sizes $\eta$ and $\kappa$ are scaled inversely to the capacity of 
the payment channel to ensure that payment channels with very large capacities do not update prices at
the same rate as channels with lower capacities. 
We also add a second-order term to the rate update which effectively drives 
faster change if the updates to the rate were in the same direction over the last two intervals. 
The step size on this term is set to $0.04$.
}

\NewPara{Metrics.}
We use the following three evaluation metrics:
(i) \emph{transaction success ratio:} the number of completed transactions
over the number of generated transactions. 
A transaction which is split at the source is complete when all of its split pieces successfully reach the destination, and
 (ii) \emph{normalized throughput:} the total amount of payments (in \eu) completed 
over the total amount of payments generated.
All of these metrics are computed over a measurement interval, set such that all algorithms are in their steady-state. 
Unless otherwise specified, we use a measurement interval of 800-1000s when running an experiment for 1010s.

\subsection{Prototype Implementation}
\label{sec:impl}

\label{sec:implementation}
To support \name, we modify
the Lightning Network Daemon (LND) \cite{lightning-implementation} which is 
currently deployed on the live Bitcoin Network.
We repurpose the router queues to queue up transactions (or HTLCs) that cannot be
immediately serviced. When a transaction spends more than 75ms in the queue,
\name marks it. The marking is echoed back via an additional field in the transaction acknowledgement (\texttt{FulfillHTLC})
 to the sender. We maintain a per-receiver state at the sender to capture
the window and number inflight on each path, as well as the queue of unattempted transactions.
Each sender finds $4$ edge-disjoint shortest paths to every destination.
We do not implement transaction-splitting.

\label{sec:validation}
We deploy our modified LND implementation \cite{lightning-implementation} 
on Amazon EC2's \texttt{c5d.4xlarge} instances with 16 CPU cores, 
16 GB of RAM, 400 GB of NVMe SSD, 
and a 10 Gbps network interface. Each instance hosts one end-host and one router.
Every LND node is run within a docker container with a dedicated bitcoin daemon \cite{bitcoind}.
We create our own regtest \cite{regtest} blockchain for the nodes. 
Channels are created 
corresponding to a scale-free graph with 10 nodes and 25 edges. 
We vary the mean channel size from 25\eu to 400\eu. 
Five circulation payment graphs are generated with each sender sending 100 tx/s (each 1\eu).
Receiving nodes communicate invoices via etcd \cite{etcd} to sending nodes
who then complete them using the appropriate scheme.
We run LND and \name on the implementation and  measure the transaction RTTs to inform propagation delays on the simulator.
We then run the same experiments on the simulator. 

\begin{figure}
    \includegraphics[width=0.8\columnwidth]{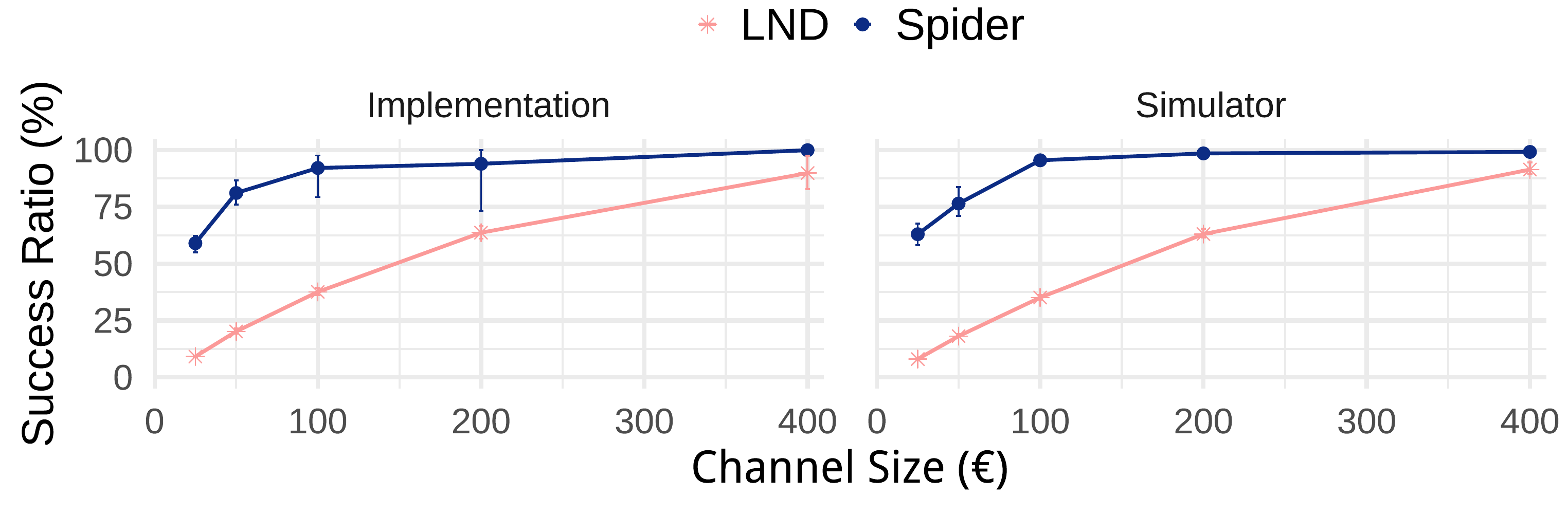}
     \caption{\small Comparison of performance on simulator and implementation for LND and \name on a 
     10 node scale-free topology with 1\eu transactions. \name outperforms LND in both settings. Further,
     the average \SR on the simulator and implementation for both schemes are within 5\% of each other.}
    \label{fig:sim impl}
\end{figure}

\Fig{sim impl} shows the average \SR that \name and LND achieve on the implementation and the simulator.
There are two takeaways: (i) \name outperforms LND in both settings and, (ii) 
the average \SR on the simulator is within 5\% of the implementation for both schemes.
Our attempts at running experiments at larger scale showed that the LND codebase 
is not optimized for high throughput. 
For example, persisting HTLC state on disk causes IO bottlenecks and variations of tens of seconds in transaction latencies
even on small topologies. 
Given the fidelity and flexibility of the simulator, we chose to use it for the remaining evaluations.



\subsection{Circulation Payment Graph Performance}
\label{sec:circulation}
\begin{figure*}[h]
    \centering
    \includegraphics[width=\textwidth]{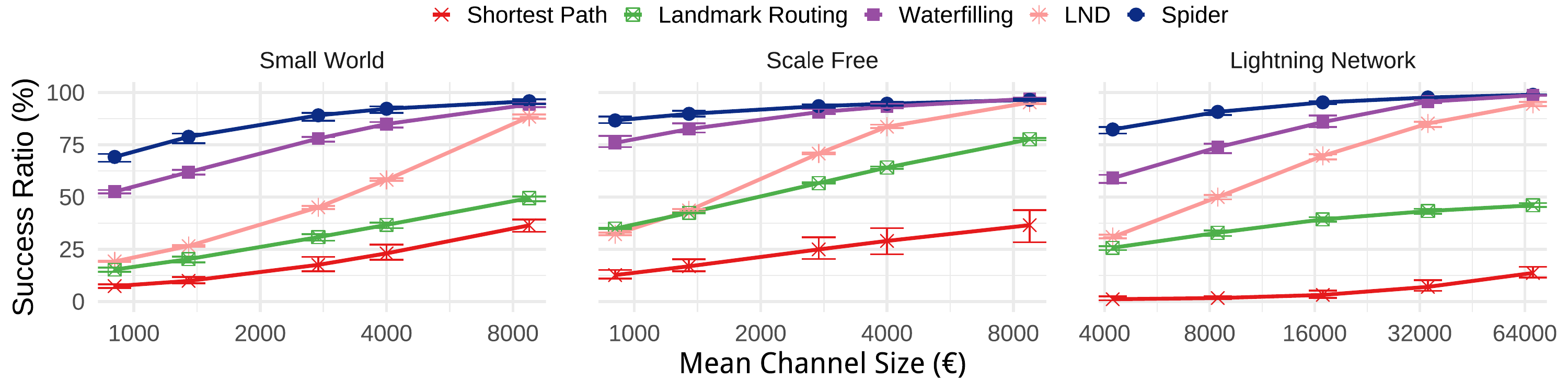}
    \caption{\small Performance of different algorithms on small-world, scale-free and Lightning Network topologies, for different 
    per sender transaction arrival rates. \name consistently outperforms all other schemes achieving near 100\%
    average \SR. Note the log scale of the x-axes.}
    \label{fig:bigexp throughput}
\end{figure*}

\begin{figure*}[t]
    \centering
    \includegraphics[width=\textwidth]{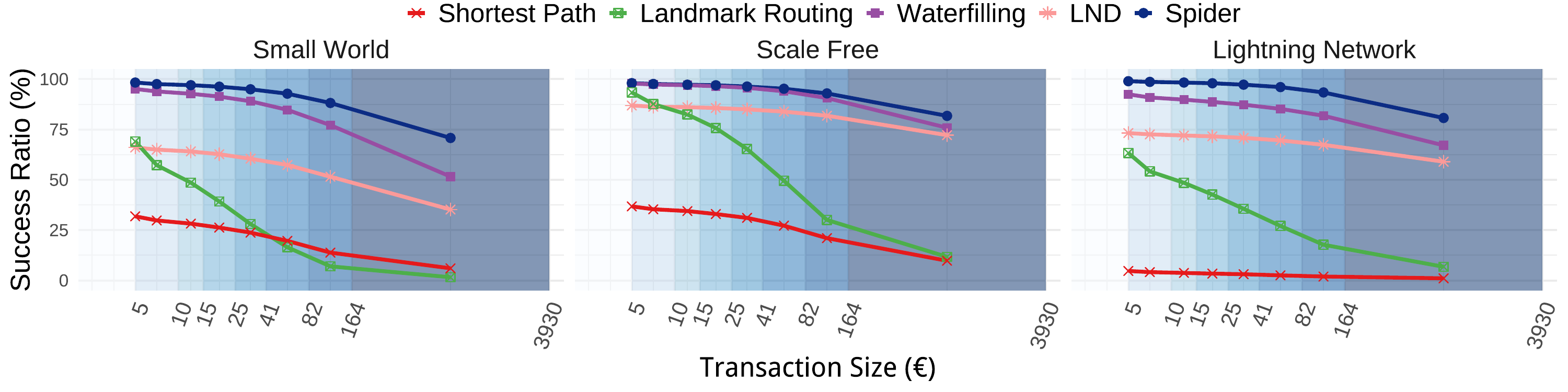}
    \caption{\small Breakdown of performance of different schemes by size of transactions completed. 
        Each point reports the \SR for transactions whose size
        belongs to the interval denoted by the shaded region. 
        Each interval corresponds roughly to a 12.5\% weight in the transaction size CDF shown in \Fig{kaggle}. The graphs correspond to the midpoints of the corresponding Lightning sampled channel sizes in \Fig{bigexp throughput}.}
    \label{fig:bigexp size}
\end{figure*}

Recall that on circulation payment graphs, \emph{all} the demand can theoretically be  routed 
if there is sufficient capacity (\S\ref{sec:throughput bounds} and \App{apx: circulation_proof}). 
However, the capacity at which a routing scheme attains 100\%
throughput depends on the  scheme's ability to balance channels: the more
balanced a scheme is, the less capacity it needs for high throughput. 


\NewPara{Efficiency of Routing Schemes}.
We run five circulation traffic matrices on our three topologies (\S\ref{sec:simulator}).
Notice that the channel sizes are much larger on the Lightning Topology compared to the other two
due to the highly skewed nature of capacities (\Fig{lnd:topo:capacity}).
We measure \SR for the transactions across different channel sizes. 
\Fig{bigexp throughput} shows that on all topologies, \name outperforms the state-of-the-art schemes. 
\name successfully routes more than 95\% of the 
transactions with less than 25\% of the capacity required by LND. 
At lower capacities, 
\name completes 2-3$\times$ more transactions than LND. 
This is because \name maintains balance in the network by responding quickly 
to queue buildup at payment channels, thus making better use of network 
capacity. 
The explicit balance-aware scheme, Waterfilling, also routes more transactions than LND. However, when operating in low capacity regimes, where many paths are congested and have near-zero available balance, senders are unable to use just balance information to differentiate  paths. As a result, \WF's performance degrades at low capacity compared to Spider which takes into account queuing delays. 

\NewPara{Size of Successful Payments}. 
{\name}'s benefits are most pronounced at larger 
transaction sizes, where packetization and congestion control helps more transactions complete.
\Fig{bigexp size} shows \SR as a function of transaction  size.
We use mean channel
sizes of 4000\eu and 16880 \eu for the synthetic and real topologies, respectively.
Each shaded region denotes a different range of transaction sizes, each corresponding to about 12.5\% of the transactions in the workload.  
A point within a range represents the average \SR for transactions in that interval 
across 5 runs.
 \name outperforms LND across all sizes, and is able
to route 5-30\% more of the largest transactions compared to LND.

\eat{\begin{figure}
    \includegraphics[width=\columnwidth]{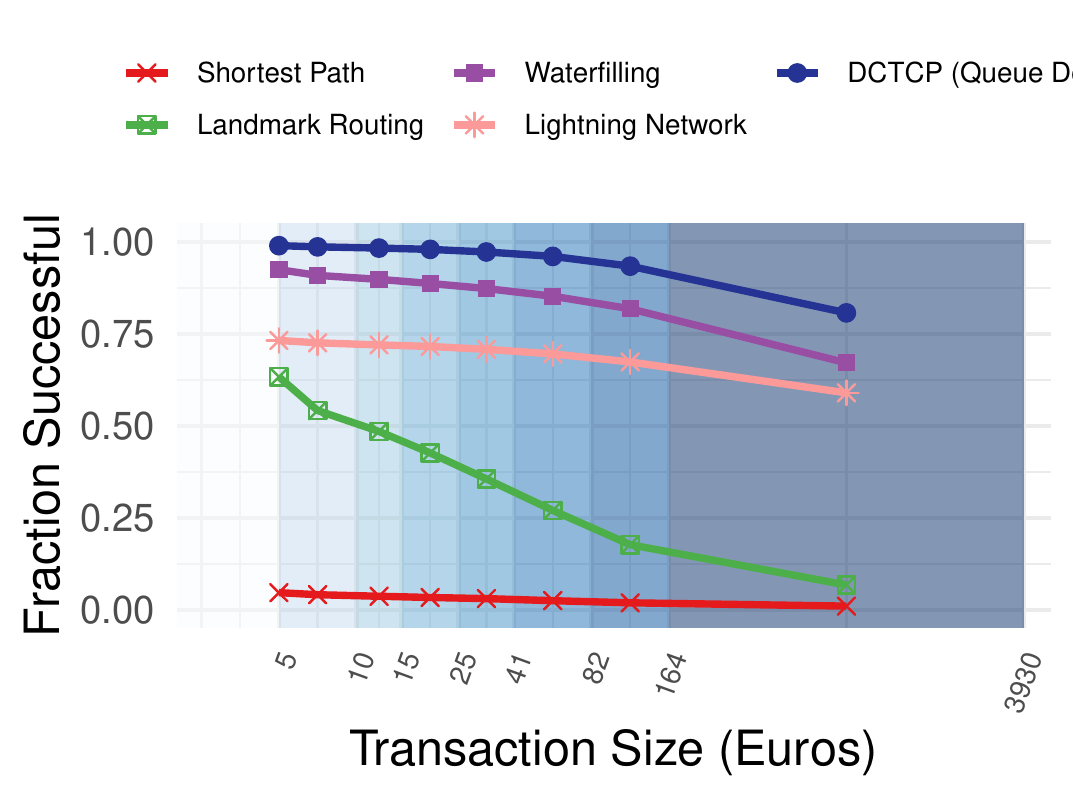}
     \caption{\small Fraction of successful payments across different sizes of attempted transactions on 
     the Lightning Network Topology with mean capacity of 26K.}
    \label{fig:lnd:succprob}

\end{figure}}

\begin{figure}
    \centering
    \includegraphics[width=0.85\columnwidth]{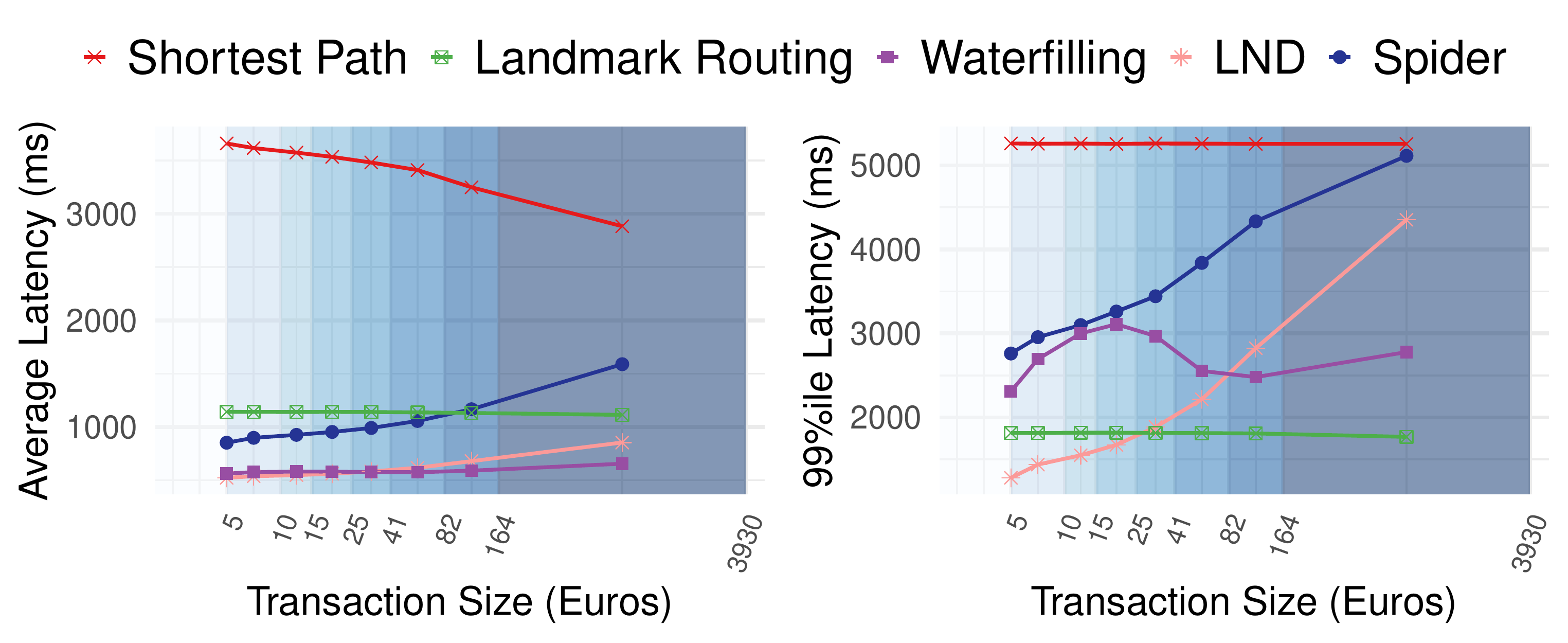}
     \caption{\small Average and 99\%ile transaction latency for different routing schemes on the Lightning topology.
     Transactions experience 1-2s of additional latency with \name relative to LND for a 20\% improvement in throughput.} 
    \label{fig:lnd comp time}
\end{figure}

\NewPara{Impact on Latency}.
We anticipate \name's rate control mechanism to increase latency.
\Fig{lnd comp time} shows the average and 99$^{th}$ percentile latency for successful transactions on the Lightning topology
as a function of transaction size. 
\name's average and tail latency increase with transaction size because larger transactions are multiplexed over longer
periods of time. 
However, the tail latency increases much more than the average because of the skew in channel sizes in the Lightning topology: 
most transactions use large channels while a few unfortunate large transactions need more time to reuse tokens from
smaller channels. Yet, the largest \name transactions experience  
at most 2 seconds of additional delay when compared to LND,
a small hit relative to the 20\% increase in overall \SR at a mean channel size of 16880\eu.
LND's latency also increases with size since it retries transactions, 
often upto 10 times until it finds a single path with enough capacity.
In contrast, \LR and Shortest path are size-agnostic in their path-choice for transactions. 

\WF pauses transactions when there is no available balance and resumes sending when balance becomes available.
Small transactions are unlikely to be paused in their lifetime while mid-size transactions are paused a few times before they complete. In contrast, large transactions are likely to be paused many times, eventually getting canceled if paused too much.
This has two implications: (i) the few large transactions that are successful with \WF are not paused much and contribute
smaller latencies than mid-size transactions, and (ii) \WF's conservative pause and send mechanism implies there is less
contention for the large transactions that are actually sent into the network, leading to smaller latencies than what 
they experience with \name.

\subsection{Effect of DAGs}
\label{sec:dag}
Real transaction demands are often not pure circulations: 
consumer nodes spend more, 
and merchant nodes receive more. 
To simulate this, we add 5 DAG payment graphs (\Sec{sec:setup}) 
to circulation payment graphs, varying the relative weight to generate 
effectively 5\%, 20\% and 40\% DAG in the total demand matrix. 
We run all schemes
on the Lightning topology with a mean channel size of 16880\eu; 
results on the synthetic topologies are in \App{app:dag}. 

\begin{figure}
    \centering
    \includegraphics[width=0.85\columnwidth]{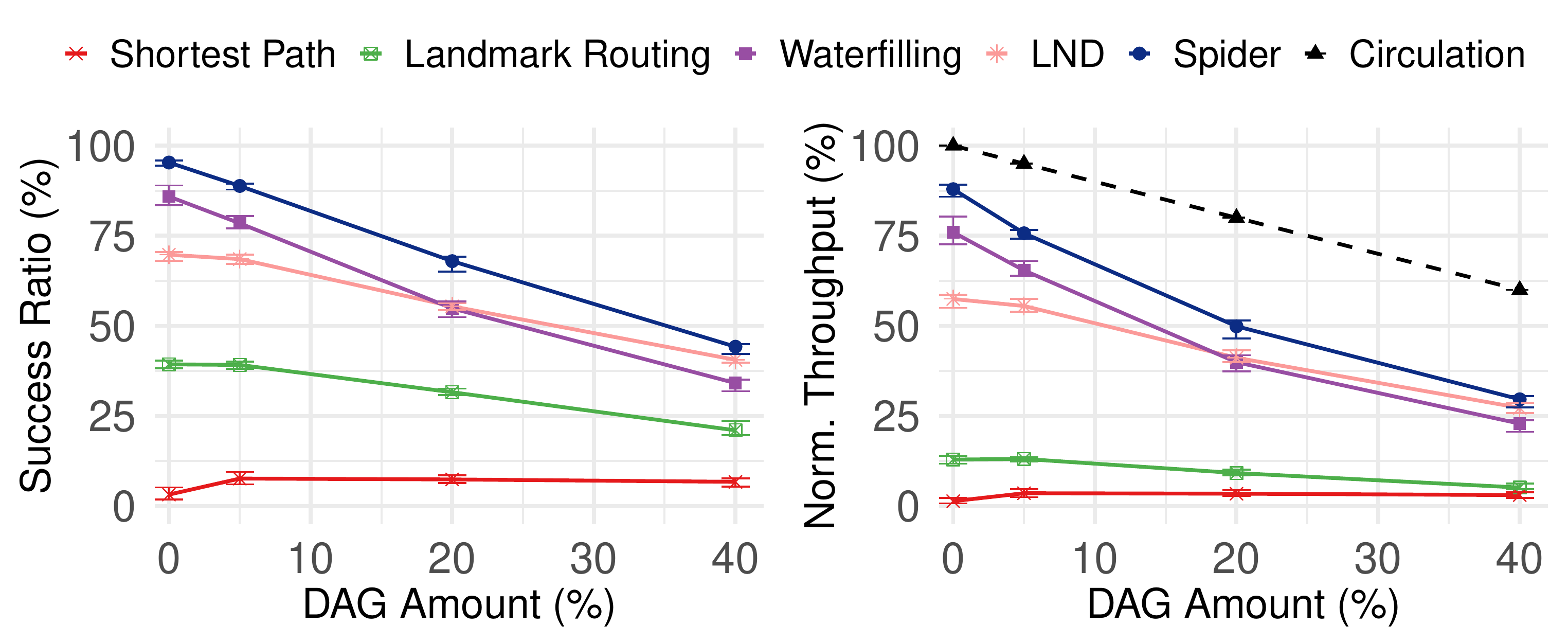}
     \caption{\small Performance of different algorithms on the Lightning topology
         as the DAG component in the transaction
     demand matrix is varied.  As the DAG amount is increased, the \NT achieved is further away 
 from the expected optimal circulation throughput.}
    \label{fig:lnd dag throughput}
\end{figure}

\Fig{lnd dag throughput} shows the \SR and \NT. 
We immediately notice that
no scheme achieves the theoretical upper bound on throughput (i.e., the \% circulation demand). 
However, throughput is closer to the bound 
when there is a smaller DAG component in the demand matrix. 
This suggests that not only is the DAG itself unroutable, it also alters the PCN balances in a way
that prevents the circulation from being fully routed. 
Further, the more DAG there is, the more affected the circulation is. 
This is because the DAG causes a deadlock (\Sec{sec:deadlock}).

To illustrate this, we run 
two scenarios: (i) a pure circulation demand $X$ for 3000s, and 
(ii) a traffic demand $(X + Y)$ containing 20\% DAG for 2000s followed by the circulation $X$ for 1000s after that.
Here, each sender sends 200\eu/s of unit-sized transactions in $X$.
We observe a time series of the \NT over the 3000s.
The mean channel size is 4000\eu and 16990\eu for the synthetic and real topologies respectively.

\begin{figure}
    \centering
    \includegraphics[width=0.85\columnwidth]{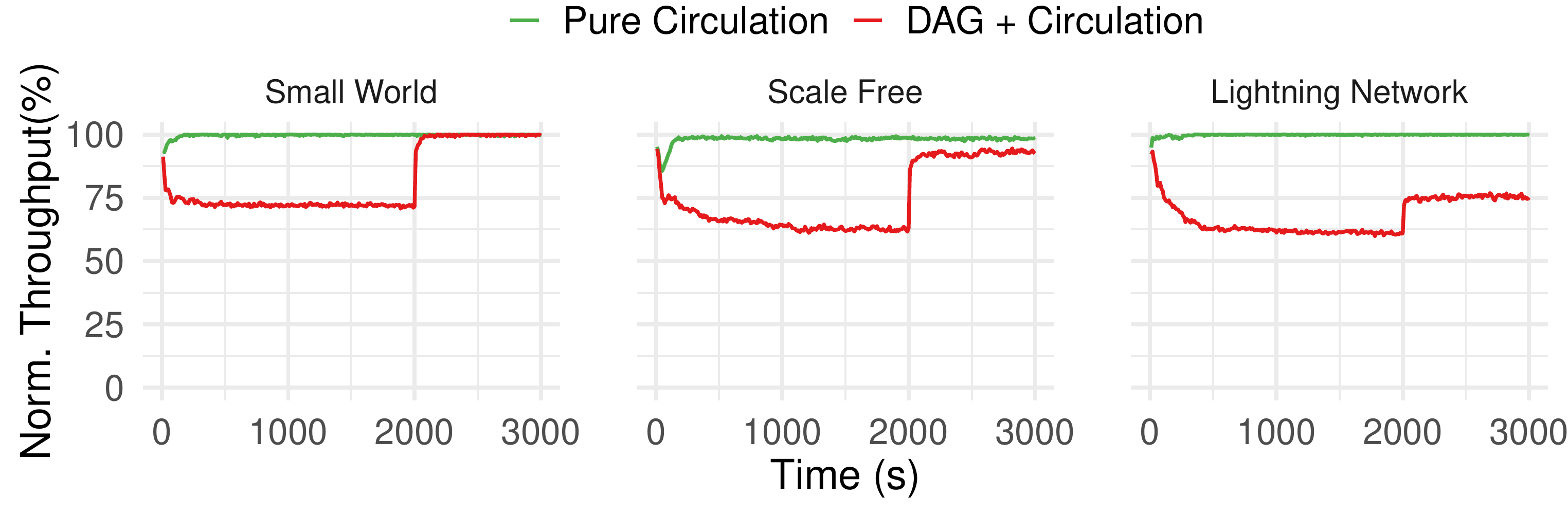}
     \caption{\small Comparing throughput when a pure circulation demand is run for 3000s to a scenario
         where a circulation demand is restored for 1000s after 2000s of a demand with 20\% DAG. The throughput
     achieved on the last 1000s of circulation is not always the expected 100\% even after the DAG is removed.}
    \label{fig:dag timeseries}

\end{figure}

\Fig{dag timeseries} shows that \name achieves 100\% throughput (normalized by the circulation demand) at steady state for the pure circulation 
demand on all topologies. 
However, when the DAG component is introduced to the demand,
it affects the topologies differently.
Firstly, we do not observe the expected 80\% throughput for the circulation in the presence
of the DAG workload suggesting that the DAG affects the circulation.
Further, even once the circulation demand is restored for the last 1000s, in the scale free
and Lightning Network topology, the throughput achieved is no longer 100\%. 
In other words, in these two topologies,
the DAG causes a deadlock that affects the circulation even after the DAG is removed.

As described in \Sec{sec:motivation}, the solution to this problem involves replenishing funds via on-chain
rebalancing,  since DAG demands continuously move money from sources to sinks. 
We therefore implement a simple rebalancing scheme where every router periodically reallocates funds between its payment channels
to equalize their \emph{available balance}. 
The frequency of rebalancing for a router, is defined by the number of successful \tus (in \eu) between consecutive rebalancing events. 
In this model, the frequency captures the on-chain rebalancing cost vs. routing fee trade-off for the router.
\begin{figure}
    \centering
    \includegraphics[width=0.85\columnwidth]{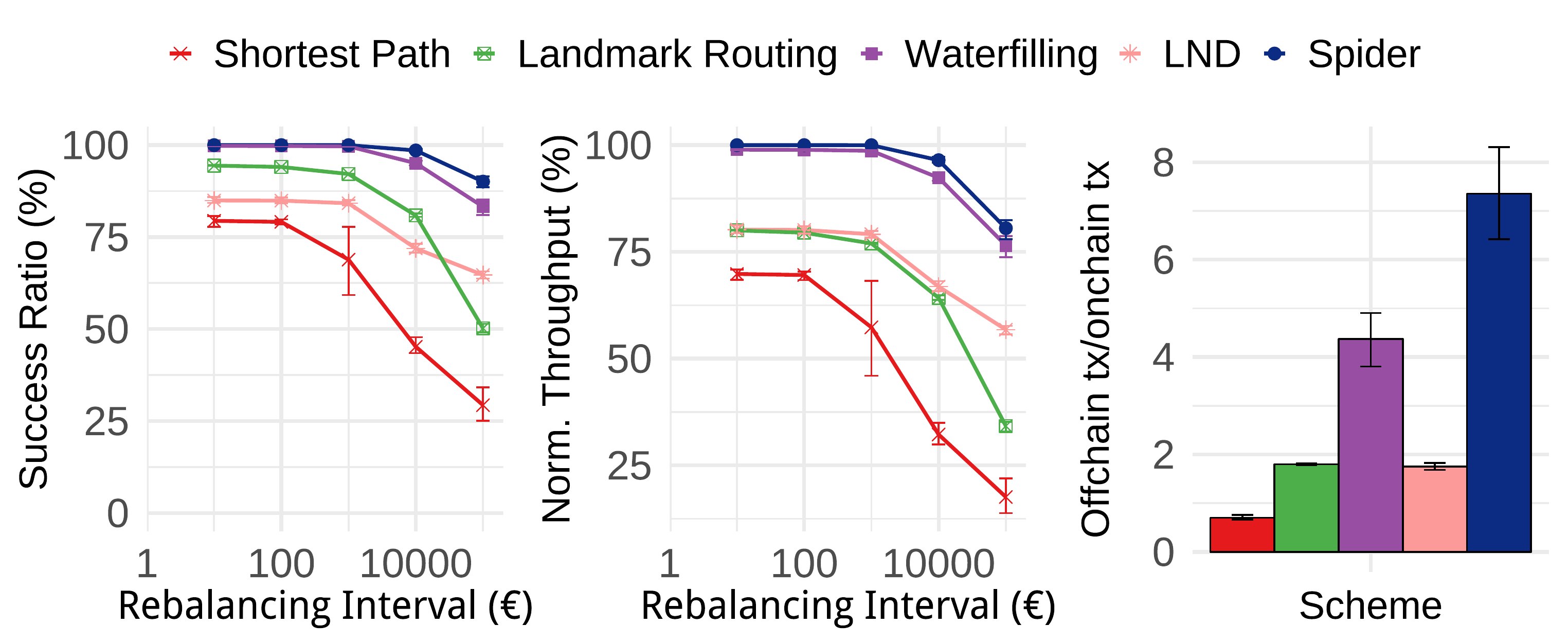}
     \caption{\small Performance of different algorithms on the Lightning topology when augmented with on-chain rebalancing.
     \name needs less frequent rebalancing to sustain high throughput.
 \name offloads 3-4x more transactions onto a PCN per blockchain transaction than LND.}
    \label{fig:lnd rebalancing}
\end{figure}

\Fig{lnd rebalancing} shows the \SR and \NT achieved by different schemes when rebalancing is enabled for the traffic
demand with 20\% DAG from \Fig{lnd dag throughput}, or \Fig{dag timeseries}. \name is able to achieve 90\% \SR even when its routers rebalance
only every 10,000\eu routed while LND is never able to sustain more than 85\% \SR even when rebalancing for every 10\eu routed. This is because LND deems a channel unusable for $5$ seconds every time a
transaction fails on it due to lack of funds and this is further worsened by its lack of transaction splitting.
This implies that when using \name, routers need to pay for only one on-chain transaction typically costing under 1\eu \cite{bitcoin-fee-chat} for every 10,000\eu routed. Thus, for a router to break even, it would have to charge 1\eu for every 10000\eu routed. This translates into significantly lower routing fees for end-users
than today's payment systems \cite{visa-fees}. 
\Fig{lnd rebalancing} also captures the same result in the form of the best offloading or number of off-chain PCN transactions per blockchain transaction achieved by each algorithm. Transactions that fail on the PCN as well as rebalancing transactions are counted towards the transactions on the blockchain. \name is able to route 7-8 times as many transactions off-chain for every blockchain transaction, a 4x improvement from the state-of-the-art LND.

\subsection{\name's Design Choices}
\label{sec:design choices}
In this section, we investigate \name's design choices with respect to the
number of paths, type of paths, and the scheduling algorithm that services \tus at \name's queues. 
We evaluate these on both the real and synthetic topologies with channel sizes
sampled from the LCSD, and scaled to have mean of 16880\eu and 4000 \eu respectively .

\begin{figure}
    \centering
    \includegraphics[width=0.5\columnwidth]{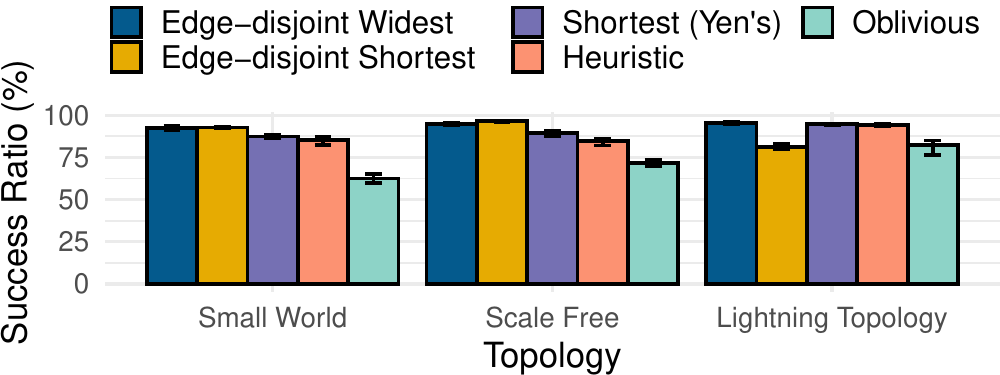}
    \caption{\small
    Performance of \name as the type of paths considered per sender-receiver pair is varied. Edge-disjoint widest
    outperforms others by 1-10\% on the Lightning Topology without being much worse on the synthetic topologies.} 
    \label{fig:path:types}
\end{figure}

\NewPara{Choice of Paths}.
We vary the type of paths that \name uses by replacing edge-disjoint widest paths with edge-disjoint
shortest paths, Yen's shortest paths \cite{yen}, oblivious paths \cite{oblivious} and a heuristic approach. 
For the widest and oblivious path computations, the channel size acts as the edge weight. 
The heuristic picks 4 paths for each flow with the highest bottleneck balance/RTT value.
\Fig{path:types} shows that edge-disjoint widest paths outperforms other approaches by 1-10\% on the Lightning Topology while being only 1-2\% worse that edge-disjoint shortest paths on the synthetic topologies.
This is because widest paths are able to utilize the capacity of the network better when there
is a large skew (\Fig{lnd:topo:capacity}) in payment channel sizes. 

\eat{shows that edge-disjoint shortest paths outperforms all other path-choices for a given demand.
Edge-disjoint paths implies that paths do not share any common capacity bottlenecks.
Choosing shorter paths also means that on average, the feedback delay from prices to rates is lower. 
Together, this enables higher \NT and \SR for transactions in the system. 
Oblivious routing, for this topology, performs poorly because on average, its path lengths are 
25\% higher than other schemes leading to longer feedback delays. 
The average bottleneck is also 30-50\% lower compared to other schemes implying that shared bottlenecks across paths could quickly deplete funds on certain paths leaving them unusable. 
}

\begin{figure}
    \centering
        \includegraphics[width=0.5\columnwidth]{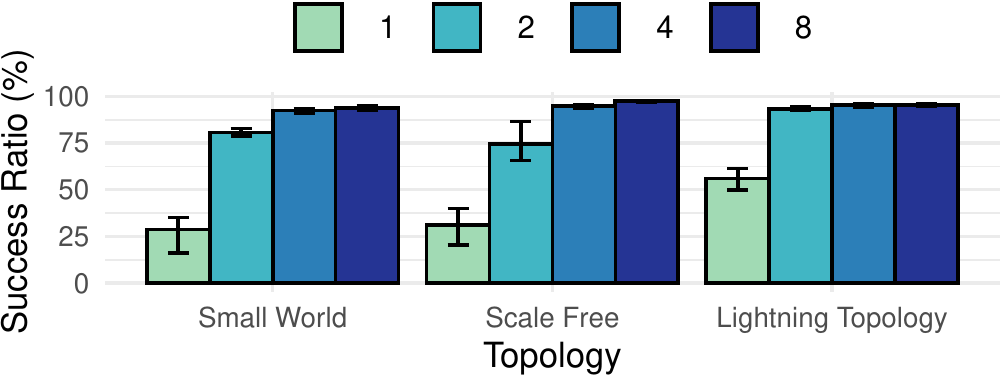}
        \caption{\small Performance of \name as the number of edge-disjoint widest paths considered 
        per sender-receiver pair is varied on different topologies. Increasing the number of 
        paths increases \SR, but the gains are low in going from 4 to 8 paths.}
   	\label{fig:num:paths} 
\end{figure}

\NewPara{Number of Paths}.
We vary the maximum number of edge-disjoint widest paths \name allows from 1 to 8.
\Fig{num:paths} shows that, as expected, the \SR increases with an increase in number of paths,
as more paths allow \name to better utilize the capacity of the \PCN. 
While moving from 1 to 2 paths results in 30-50\% improvement in \SR, moving from 4 to 8 paths 
has negligible benefits (<5\%). 
This is because the sparseness of the three \PCN topologies causes most flows to have at most 5-6 edge-disjoint 
widest paths.
Further, \name prefers paths with smaller RTTs since they receive feedback faster 
resulting in the shortest paths contributing most to the overall rate for the flow. 
As a result, we use 4 paths for \name.

\begin{figure}
    \centering
    \includegraphics[width=0.5\columnwidth]{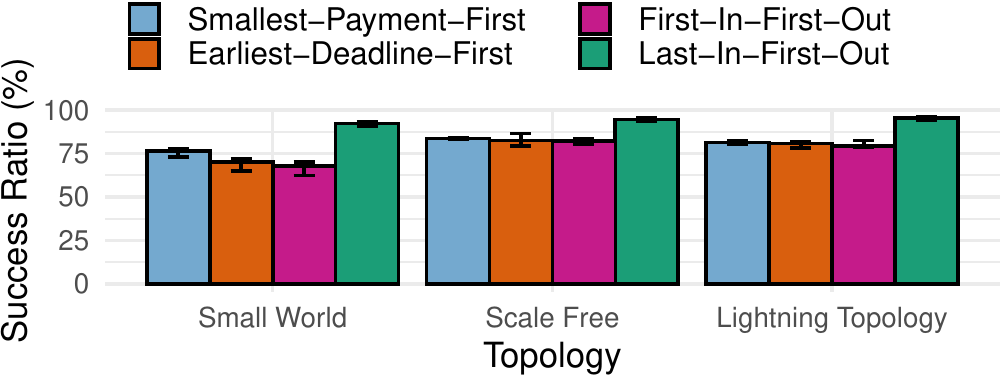}
    \caption{\small
    Performance of \name as the scheduling algorithm at the sender and router queues is varied. Last in first out
    outperforms all other approaches by over 10\% on all topologies.}
    \label{fig:scheduling:alg}
\end{figure}

\NewPara{Scheduling Algorithms}.
We modify the scheduling algorithm at the per-destination queues at the sender as well as the router
queues in \name to process transactions as per First-In-First-Out (FIFO), Earliest-Deadline-First (EDF) 
and Smallest-Payment-First (SPF) in addition to the LIFO baseline.
\Fig{scheduling:alg} shows that LIFO achieves a \SR that is 10-28\% higher than its counterparts.
This is because LIFO prioritizes transactions that are newest or furthest from their deadlines
and thus, most likely complete especially when the PCNs is overloaded.
\name's rate control results in long wait times in the sender queues themselves. This causes FIFO and EDF that 
send out transactions closest to their deadlines to time out immediately in the network resulting in 
poor throughput. 
When SPF deprioritizes large payments at router queues, they 
consume funds from other payment channels for longer, reducing the effective capacity of the network.

\eat{
\NewPara{Sensitivity to queue delay threshold}

We now measure how sensitive \name is to the queue delay threshold at which we start marking packets. 
We fix the demand and capacity on a given topology and vary the threshold at which we mark packets from 15ms to 600 ms. We notice 
\begin{figure}
    \begin{subfigure}[t]{0.5\columnwidth}
    	\includegraphics[width=\columnwidth]{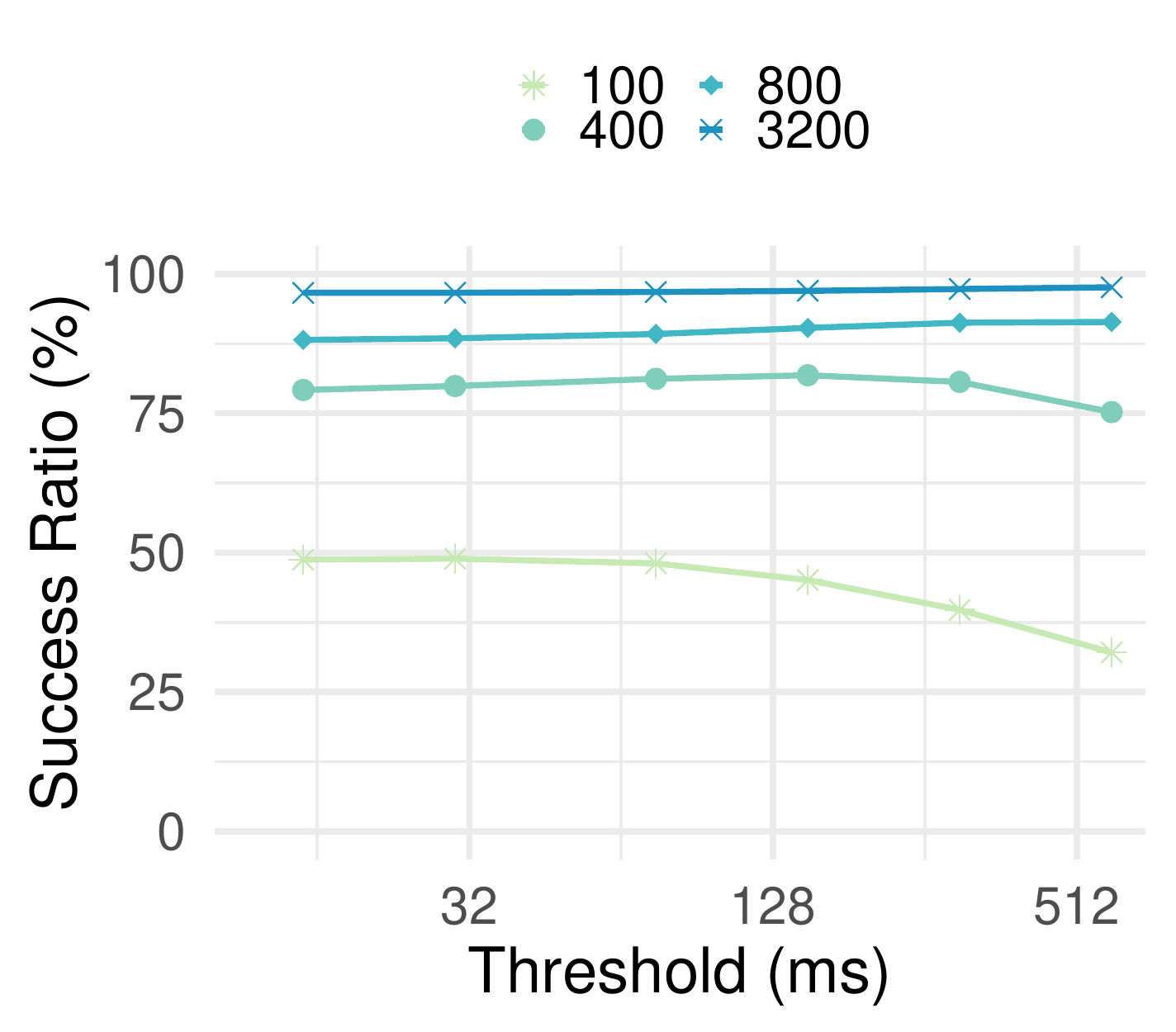}
        \caption{Small World Topology (Uniform)}
   	\label{fig:sw:topo:qthreshold}
    \end{subfigure}~
    \begin{subfigure}[t]{0.5\columnwidth}
    	\includegraphics[width=\columnwidth]{figures/sw_50_routers_uniform_credit_DCTCPQ_thresholdSuccRatio.pdf}
        \caption{Lightning Network (temporary)}
   	\label{fig:lnd:topo:qthreshold}
    \end{subfigure}
    \caption{\small Sensitivity to queue delay threshold across different mean capacities and topologies}
    \label{fig:qthreshold}
\end{figure}
}

\subsection{Additional Results}
\label{sec:additional results}
In addition to the results described so far, we run additional experiments that are described in the Appendices.

\begin{enumerate}[noitemsep,topsep=0pt,parsep=0pt,partopsep=0pt]
\item We compare \name to Celer, as proposed in a white-paper~\cite{celer}, and show that \name outperforms Celer's \SR by 2x 
on a scale free topology with 10 nodes and 25 edges (\App{app:celer}).
\item We evaluate the schemes on the synthetic and real topologies with a simpler channel size 
    distribution where all channels have equal numbers of tokens. Even in this scenario,
    \name is able to successfully route more than 95\% of the transactions with 
    less than 25\% of the capacity required by LND (\App{app:uniform csd}).
\item We evaluate the schemes for their fairness across multiple payments and show that \name does not
hurt small payments to gain on throughput (\App{app:fairness}).
\item We show the effect of DAG workloads on synthetic topologies. In particular, we identify deadlocks  
with those topologies too and show that \name requires rebalancing only every 10,000\eu successfully routed
to sustain high \SR and \NT (\App{app:dag}).
\end{enumerate}

\section{Related Work} 
\label{sec:related}

\NewPara{\PCN Improvements}. Nodes in current Lightning Network implementations, 
maintain a local view of the network topology and source-route transactions along the shortest path \cite{lightning-implementation, c-lightning}.
Classical max-flow-based alternatives are impractical for the Lightning Network that has over $5000$ nodes and $30,000$ channels \cite{1ml-vis, blockchain-caffe} due to their computational complexity. 
Recent proposals have used a modified version of max-flow that differentiates based on the size of transactions \cite{flash}. However, inferring the size of payments is hard in an onion-routed network like Lightning.

Two main alternatives to max-flow routing have been proposed: landmark routing and embedding-based routing.
In \emph{landmark routing}, select routers (landmarks) store routing tables for the rest of the network, and nodes only route transactions to a landmark \cite{tsuchiya1988landmark}. This approach is used in Flare \cite{prihodko2016flare} and 
\SW \cite{malavolta2016silentwhispers,moreno2015privacy}.
\emph{Embedding-based} or \emph{distance-based} routing learns a vector embedding for each node, such that nodes that are close in network hop distance are also close in embedded space.
Each node relays each transaction to the neighbor whose embedding is closest to the destination's embedding.
VOUTE \cite{voute} and \SM \cite{speedymurmurs} use embedding-based routing.
Computing and updating the embedding dynamically as the topology and link balances change is a primary challenge of these approaches. Our experiments and prior work \cite{spiderhotnets} show that \name outperforms both approaches.

PCN improvements outside of the routing layer focus on rebalancing existing payment channels more easily \cite{channel-factories, revive}. 
Revive \cite{revive} leverages cycles within channels wanting to rebalance and initiates balancing off-chain payments 
between them. These techniques are complementary to \name and can be used to enhance overall performance. However, \Sec{sec:dag} shows that a more general rebalancing scheme that moves funds 
at each router independently fails to achieve high throughput without a balanced routing scheme.





\NewPara{Utility Maximization and Congestion Control}.
Network Utility Maximization (NUM) is a popular framework for developing
decentralized transport protocols in data networks
to optimize a fairness objective \cite{kelly2005stability}. 
NUM uses link ``prices'' derived from the solution
 to the utility maximization problem, and senders compute rates based on these router 
prices. 
Congestion control algorithms that use
buffer sizes or queuing delays as router signals \cite{RCP,VCP,dctcp} are closely related. 
While the Internet congestion control literature has focused on links with fairly stable
capacities, this paper shows that they can be effective even in networks
with capacities dependent on the input rates themselves.
Such problems have also been explored in the context of ride-sharing, for instance \cite{banerjee2015pricing,banerjee2016dynamic}, and require new innovation in both formulating and solving routing problems.

\newcommand{\limitsec}{Conclusion}
\section{\limitsec}
\label{sec:limit}

We motivate the need for efficient routing on {\PCN}s and 
propose \name, a protocol for balanced, high-throughput routing in PCNs.
\name uses a packet-switched architecture, multi-path congestion control, and
and in-network scheduling. \name achieves nearly 100\% 
throughput on circulation payment demands across both synthetic and real topologies.
We show how the presence of DAG payments causes deadlocks
that degrades circulation throughput, necessitating on-chain intervention. 
In such scenarios, \name is able to support 4x more transactions than the state-of-the-art on the PCN itself.

This work shows that \name needs less on-chain rebalancing to relieve deadlocked PCNs.
However, it remains to be seen if deadlocks can be prevented altogether.
\name relies on routers signaling queue buildup correctly to the senders, but this work
does not analyze incentive compatibility for rogue routers
aiming to maximize fees. 
A more rigorous treatment of the privacy 
implications of \name routers relaying queuing delay is left to future work.




\section{Acknowledgements}
\label{sec:ack}

We thank Andrew Miller, Thaddeus Dryja, Evan Schwartz, Vikram Nathan, and Aditya Akella for their detailed feedback. We also thank the Sponsors of Fintech@CSAIL, the Initiative for CryptoCurrencies and Contracts (IC3), Distributed Technologies Research Foundation, Input-Output Hong Kong Inc, the CISCO Research Center, the National Science Foundation under grants CNS-1718270, CNS-1563826, CNS-1910676, CCF-1705007 and CNS-1617702, and the Army Research Office under grant W911NF1810332 for their support.

\newpage
\bibliographystyle{abbrv} 
\begin{small}
\bibliography{nsdi20}
\end{small}

\newpage
\appendix
\noindent
\textbf{\Large Appendices} 

\section{Circulations and Throughput Bounds} 
\label{apx: circulation_proof}
For a network $G(V,E)$ with set of routers $V$, we define a {\em payment graph} $H(V,E_H)$ as a graph that specifies the payment demands between different users.
The weight of any edge $(i,j)$ in the payment graph is the average rate at which user $i$ seeks to transfer funds to user $j$. 
A {\em circulation graph} $C(V,E_C)$ of a payment graph is any subgraph of the payment graph in which the weight of an edge $(i,j)$ is at most the weight of $(i,j)$ in the payment graph, and moreover the total weight of incoming edges is equal to the total weight of outgoing edges for each node. 
Of particular interest are {\em maximum circulation graphs} which are circulation graphs that have the highest total demand (i.e., sum of edge weights), among all possible circulation graphs.
A maximum circulation graph is not necessarily unique for a given payment graph. 

\begin{prop} \label{prop: bounds}
Consider a payment graph $H$ with a maximum circulation graph $C^*$.
Let $\nu(C^*)$ denote the total demand in $C^*$.
Then, on a network in which each payment channel has at least $\nu(C^*)$ units of escrowed funds, there exists a balanced routing scheme that can achieve a total throughput of $\nu(C^*)$. 
However, no balanced routing scheme can achieve a throughput greater than $\nu(C^*)$ on any network. 
\end{prop}

\begin{figure}
    \centering
    \begin{tabular}{c c c}
        \begin{subfigure}{0.3\columnwidth}
            \centering
            \includegraphics[width=\columnwidth]{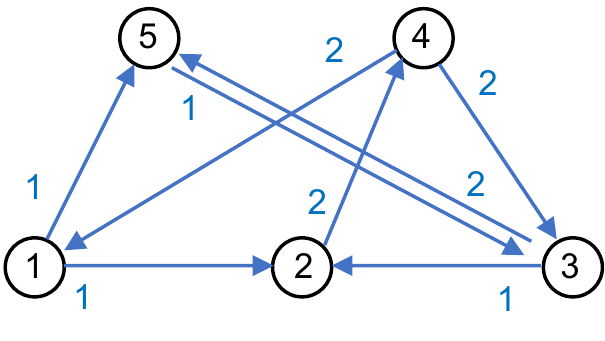}
            \subcaption{Payment graph}
\label{fig:app_payment_graph_1}
        \end{subfigure} &
        \begin{subfigure}{0.3\columnwidth}
            \centering
            \includegraphics[width=\columnwidth]{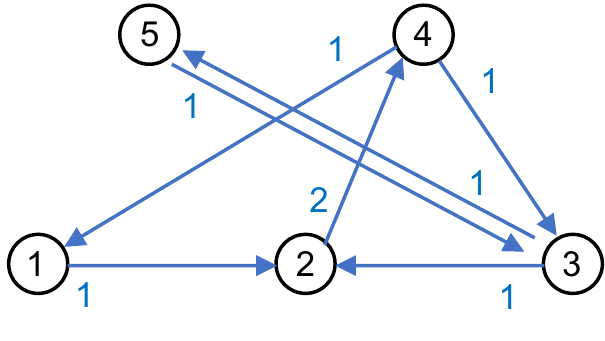}
            \subcaption{Circulation}
\label{fig:app_payment_graph_2}
        \end{subfigure} &
        \begin{subfigure}{0.3\columnwidth}
            \centering
            \includegraphics[width=\columnwidth]{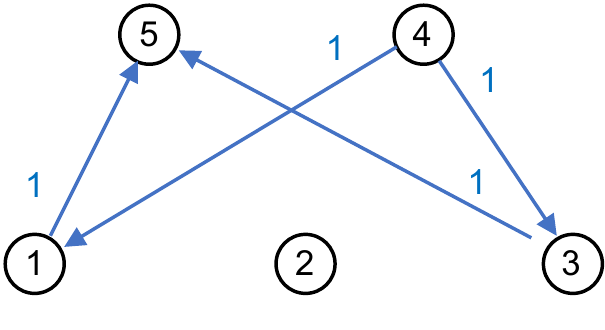}
            \subcaption{DAG}
\label{fig:app_payment_graph_3}
        \end{subfigure}         
    \end{tabular}
    \caption{\small Example payment graph (denoted by blue lines) for a five node network (left). It decomposes into a maximum circulation and DAG components as shown in (b) and (c).}
\label{fig:app_payment_graph}
\end{figure}

\begin{proof}
Let $w_{C^*}(i,j)$ denote the payment demand from any user $i$ to user $j$ in the maximum circulation graph $C^*$.
To see that a throughput of $\nu(C^*)$ is achievable, consider routing the circulation demand along the shortest paths of any spanning tree $T$ of the payment network $G$.
In this routing, for any pair of nodes $i, j \in V$ there exists a unique path from $i$ to $j$ in $T$ through which $w_{C^*}(i, j)$ amount of flow is routed.
We claim that such a routing scheme is perfectly balanced on all the links.
This is because for any partition $S, V \backslash S$ of $C^*$, the net flow going from $S$ to $V \backslash S$ is equal to the net flow going from $V \backslash S$ to $S$ in $C^*$.
Since the flows along an edge $e$ of $T$ correspond precisely to the net flows across the partitions obtained by removing $e$ in $T$ , it follows that the flows on $e$ are balanced as well.
Also, for any flow $(i,j)$ in the demand graph $C^*$, the shortest path route from $i$ to $j$ in $T$ can cross an edge $e$ at most once. 
Therefore the total amount of flow going through an edge is at most the total amount of flow in $C^*$, which is $\nu(C^*)$. 

Next, to see that no balanced routing scheme can achieve a throughput greater than $\nu(C^*)$, assume the contrary and suppose there exists a balanced routing scheme $\mathtt{SCH}$ with a throughput greater than $\nu(C^*)$.
Let $H_{\mathtt{SCH}} \subseteq H$ be a payment graph where the edges represent the portion of 
demand that is actually routed in $\mathtt{SCH}$. 
Since $\nu(H_\mathtt{SCH}) > \nu(C^*)$, $H_\mathtt{SCH}$ is not a circulation and there exists a partition $S, V \backslash S$ such that the net flow from $S$ to $V \backslash S$ is strictly greater than the net flow from $V \backslash S$ to $S$ in $H_\mathtt{SCH}$.
However, the net flows routed by $\mathtt{SCH}$ across the same partition $S, V \backslash S$ in $G$ are balanced (by assumption) resulting in a contradiction.
Thus we conclude there does not exist any balanced routing scheme that can achieve a throughput greater than $\nu(C^*)$. 
\end{proof}

\section{Optimality of \name}
\label{apx:fluidmodel}
\begin{figure}[t]
\begin{center}
\includegraphics[width=3.3in]{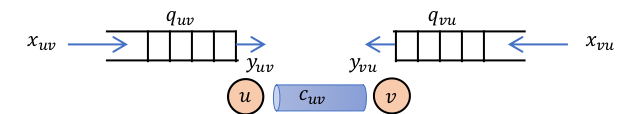}
\caption{\small Model of queues at a payment channel between nodes $u$ and $v$. $x_{uv}$ and $y_{uv}$ denote the rates at 
which \tus for $v$ arrive into and get serviced at the queue at $u$ respectively. $c_{uv}$ is the capacity of 
the payment channel and $q_{uv}$ denotes the total number of \tus waiting in $u$'s queue to be serviced.}
\label{fig:queue model}
\end{center}
\end{figure}

\subsection{Fluid Model}
\label{apx:fluid model}
In this section we describe a fluid model approximation of the system dynamics under \name's protocol. 
Following a similar notation as in \S\ref{sec:model}, for a path $p$ we let $x_p(t)$ denote the rate of flow on it at time $t$. 
For a channel $(u,v)$ and time $t$, let $q_{u,v}(t)$ be the size of the queue at router $u$, $f_{u,v}(t)$ be the fraction of incoming packets that are marked at $u$, $x_{u,v}(t)$ be the total rate of incoming flow at $u$, and $y_{u,v}(t)$ be the rate at which transactions are serviced (\ie forwarded to router $v$) at $u$. All variables are real-valued.
We approximate \name's dynamics via the following system of equations
\begin{align}
\dot{x}_p(t) &= \left[ \frac{x_p(t)}{\sum_{p'\in\mathcal{P}_{i_p, j_p}} x_{p'}(t)} - \sum_{(u,v) \in p} f_{u,v}(t) x_p(t) \right]_{x_p(t)}^+ \forall p \in \mathcal{P} \label{eq:xp update} \\
\dot{q}_{u,v}(t) &= [x_{u,v}(t) - y_{u,v}(t)]^+_{q_{u,v}(t)} \quad \forall (u,v) \in E \label{eq:q update}\\
\dot{f}_{u,v}(t) &= [q_{u,v}(t) - q_\mathrm{thresh}]_{f_{u,v}(t)}^+  \quad \forall (u,v) \in E, \label{eq:f update}
\end{align}
where $y_{u,v}(t) = y_{v,u}(t) =$
\begin{align}
\begin{cases*}
      \frac{c_{u,v}}{2\Delta} & if $q_{u,v}(t) > 0 \ \& \ q_{v,u}(t) > 0 $ \\
      \min \{\frac{c_{u,v}}{2\Delta}, x_{v,u}(t)\} & if $q_{u,v}(t) > 0 \ \& \ q_{v,u}(t) = 0 $ \\
      \min\{\frac{c_{u,v}}{2\Delta}, x_{u,v}(t)\} & if $q_{u,v}(t) = 0 \ \& \ q_{v,u}(t) > 0 $ \\
      \min\{\frac{c_{u,v}}{2\Delta}, x_{u,v}(t), x_{v,u}(t)\} & if $q_{u,v}(t) = 0 \ \& \ q_{v,u}(t) = 0 $ \\
\end{cases*} \label{eq:y update}
\end{align}
for each $(u,v) \in E$. Let $i_p$ and $j_p$ denote the source and destination nodes for path $p$ respectively. Then,
$\mathcal{P}_{i_p,j_p}$ denotes the set of all paths $i_p$ uses to route to $j_p$. 
Equation~\eqref{eq:xp update} models how the rate on a path $p$ increases upon receiving successful acknowledgements or decreases if the packets are marked, per Equations~\eqref{eq:window dec} and~\eqref{eq:window inc} in \S\ref{sec:transport layer endhosts}.
If the fraction of packets marked at each router is small, then the aggregate fraction of packets that return marked on a path $p$ can be approximated by the sum $\sum_{(u,v)\in p} f_{u,v}$~\cite{srikant2012mathematics}. 
Hence the rate which marked packets arrive for a path $p$ 
is $\sum_{(u,v)\in p} f_{u,v} x_p$.
Similarly, the rate which successful acknowledgements are received on a path $p$ is $x_p(1 - \sum_{(u,v)\in p} f_{u,v})$, which can be approximated as simply $x_p$ if the marking fractions are small. 
Since \name increases the window by $1/(\sum_{p'\in \mathcal{P}_{i_p,j_p}} w_{p'})$ for each successful acknowledgement received, the average rate at which $x_p$ increases is $x_p/(\sum_{p'\in \mathcal{P}_{i_p,j_p}} x_{p'})$.
Lastly, the rate $x_p$ cannot become negative; so if $x_p = 0$ we disallow $\dot{x}_p$ from being negative. 
The notation $(x)^+_y$ means $x$ if $y > 0$ and $0$ if $y = 0$. 

Equations~\eqref{eq:q update} and~\eqref{eq:f update} model how the queue sizes and fraction of packets marked, respectively, evolve at the routers. 
For a router $u$ in payment channel $(u,v)$, by definition $y_{u,v}$ is the rate at which transactions are serviced from the queue $q_{u,v}$, while transactions arrive at the queue at a rate of $x_{u,v}$ (Figure~\ref{fig:queue model}). 
Hence the net rate at which $q_{u,v}$ grows is given by the difference $x_{u,v} - y_{u,v}$. 
The fraction of packets marked at a queue grows if the queue size is larger than a threshold $q_\mathrm{thresh} $, and drops otherwise, as in Equation~\eqref{eq:f update}.  
This approximates the marking model of \name (\S\ref{sec:router design}) in which packets are marked at a router if their queuing delay exceeds a threshold. 

To understand how the service rate $y_{u,v}$ evolves (Equation~\eqref{eq:y update}), we first make the approximation that the rate at which transactions are serviced from the queue at a router $u$ is equal to the rate at which tokens are replenished at the router, \ie $y_{u,v} = y_{v,u}$ for all $(u,v)\in E$. 
The precise value for $y_{u,v}$ at any time, depends on both the arrival rates and current occupancy of the queues at routers $u$ and $v$. 
If both $q_{u,v}$ and $q_{v,u}$ are non-empty, then there are no surplus of tokens available within the channel. 
A token when forwarded by a router is unavailable for $\Delta$ time units, until its acknowledgement is received. 
Therefore the maximum rate at which tokens on the channel can be forwarded is $c_{u,v}/\Delta$, implying $y_{u,v} + y_{v,u} = c_{u,v}$ or $y_{u,v} = y_{v,u} = c_{u,v}/(2\Delta)$ in this case. 
If $q_{u,v}$ is non-empty and $q_{v,u}$ is empty, then there are no surplus tokens available at $u$'s end. 
Router $v$ however may have tokens available, and service transactions at the same rate at which they are arriving, \ie $y_{v,u} = x_{v,u}$. 
This implies tokens become available at router $u$ at a rate of $x_{v,u}$ and hence $y_{u,v} = x_{v,u}$. 
However, if the transaction arrival rate $x_{v,u}$ is too large at $v$, it cannot service them at a rate more than $c_{u,v}/(2\Delta)$ and a queue would start building up at $q_{v,u}$. 
The case where $q_{u,v}$ is empty and $q_{v,u}$ is non-empty follows by interchanging the variables $u$ and $v$ in the description above. 
Lastly, if both $q_{u,v}$ and $q_{v,u}$ are empty, then the service rate $y_{u,v}$ can at most be equal to the arrival rate $x_{v,u}$. 
Similarly $y_{v,u}$ can be at most $x_{u,v}$. 
Since $y_{u,v} = y_{v,u}$ by our approximation, we get the expression in Equation~\eqref{eq:y update}. 

We have not explicitly modeled delays, and have made simplifying approximations in the fluid model above.
Nevertheless this model is useful for gaining intuition about the first-order behavior of the \name protocol. 
In the following section, we use this model to show that \name finds optimal rate allocations for a parallel network topology. 

\subsection{Proof of Optimality}

\begin{figure}[t]
\begin{center}
\includegraphics[width=3.3in]{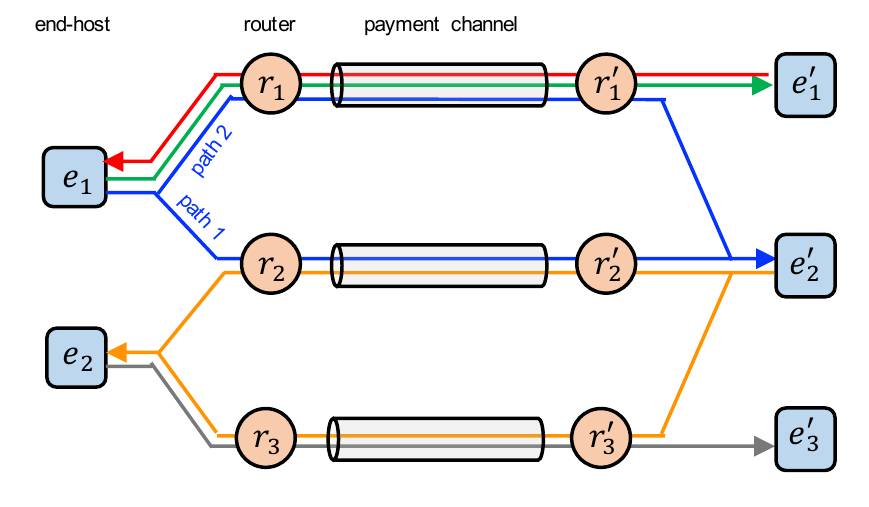}
\caption{\small Example of a parallel network topology with bidirectional flows on each payment channel.
}
\label{fig:mptcp_proof}
\end{center}
\vspace*{-3mm}
\end{figure}

Consider a PCN comprising of two sets of end-hosts $\{e_1,\ldots,e_m\}$ and $\{e'_1,\ldots,e'_n\}$ that are connected via $k$ parallel payment channels $(r_1, r'_1), \ldots, (r_k, r'_k)$ as shown in Figure~\ref{fig:mptcp_proof}. 
The end-hosts from each set have demands to end-hosts on the other set. 
The end-hosts within a set, however, do not have any demands between them. 
Let the paths for different source-destination pairs be such that for each path $p$, if $p$ contains a directed edge $(r_i, r'_i)$ for some $i$ then there exists another path (for a different source-destination pair) that contains the edge $(r'_i, r_i)$. 
We will show that running \name on this network results in rate allocations that are an optimal solution to the optimization problem in Equations~\eqref{eq:LP}--\eqref{eq:LP end}. 
Under a fluid model for \name as discussed in \S\ref{apx:fluid model}, assuming convergence, we observe that in the steady-state the time derivatives of the rate of flow of each path (Equation~\eqref{eq:xp update}) must be non-positive, \ie
\begin{align}
 \frac{1}{\sum_{p'\in\mathcal{P}_{i_p,j_p}} x_{p'}^*} - \sum_{(u,v) \in p} f_{u,v}^*  ~~  
 \begin{cases*}
      =0 & if $x_p^* > 0$ \\
      \leq 0 & if $x_p^* = 0$ \\
\end{cases*} 
 \quad \forall p \in \mathcal{P}, \label{eq:x balance}
\end{align} 
where the superscript $^*$ denotes values at convergence (\eg $x_p^*$ is the rate of flow on path $p$ at convergence). 
Similarly, the rate of growth of the queues must be non-positive, or 
\begin{align}
x_{u,v}^* ~~  \begin{cases*}
      = y_{u,v}^* & if $q_{u,v}^* > 0$ \\
      \leq y_{u,v}^* & if $q_{u,v}^* = 0$ \\
\end{cases*} 
\quad \forall (u,v) \in E. \label{eq:queue balance}
\end{align}
Now, consider the optimization problem in Equations~\eqref{eq:LP}--\eqref{eq:LP end} for this parallel network. 
For simplicity we will assume the sender-receiver demands are not constrained. 
From Equation~\eqref{eq:queue balance} above, the transaction arrival rates $x_{u,v}^*$ and  $x_{v,u}^*$ for a channel $(u,v)$ satisfy the capacity constraints in Equation~\eqref{eq:LP capacity}. 
This is because $x_{u,v}^* \leq y_{u,v}^*$ from Equation~\eqref{eq:queue balance} and $y_{u,v}(t)$ is at most $\frac{c_{u,v}}{2\Delta}$ from Equation~\eqref{eq:y update}. 
Similarly the transaction arrival rates also satisfy the balance constraints in Equation~\eqref{eq:LP balance}. 
To see this, we first note the that the queues on all payment channels through which a path (corresponding to a sender-receiver pair) passes must be non-empty. 
For otherwise, if a queue $q_{u,v}^*$ is empty then the fraction of marked packets on a path $p$ through $(u,v)$ goes to 0, and the rate of flow $x_p^*$  would increase as per Equation~\eqref{eq:xp update}. 
Therefore we have $x_{u,v}^* = y_{u,v}^*$ (from Equation~\eqref{eq:queue balance}) for every channel. 
Combining this with $y_{u,v}(t) = y_{v,u}(t)$ (Equation~\eqref{eq:y update}), we conclude that the arrival rates are balanced on all channels. 
Thus the equilibrium rates $\{x_p^*: p\in\mathcal{P}\}$ resulting from \name are in the feasible set for the routing optimization problem. 

Next, let $\lambda_{u,v} \geq 0$ and $\mu_{u,v} \in \mathbb{R}$ be the dual variables corresponding to the capacity and balance constraints, respectively, for a channel $(u,v)$. 
Consider the following mapping from $f_{u,v}^*$ to $\lambda_{u,v}$ and $\mu_{u,v}$
\begin{align}
\lambda_{u,v}^* &\leftarrow (f_{u,v}^* + f_{v,u}^*)/2 \quad \forall (u,v)\in E \label{eq:lambda map} \\
\mu_{u,v}^* &\leftarrow f_{u,v}^*/2 \quad \forall (u,v) \in E, \label{eq:mu map}
\end{align}
where the superscript $^*$ on the dual variables indicate that they have been derived from the equilibrium states of the \name protocol. 
Since $f_{u,v}(t)$ is always non-negative (Equation~\eqref{eq:f update}), we see that $\lambda_{u,v}^* \geq 0$ for all $(u,v)$. 
Therefore $\{\lambda_{u,v}^*: (u,v) \in E\}$ and $\{\mu_{u,v}^*: (u,v) \in E\}$ are in the feasible set of the dual of the routing optimization problem. 

Next, we have argued previously that the queues on all payment channels through which a path (corresponding to a sender-received pair) passes must be non-empty. 
While we used this observation to show that the channel rates $x_{u,v}^*$ are balanced, it also implies that the rates are at capacity, \ie $x_{u,v}^* = c_{u,v}/(2\Delta)$, or $x_{u,v}^* + x_{v,u}^* = c_{u,v}/\Delta$ for all $(u,v)$. 
This directly follows from Equation~\eqref{eq:queue balance} and the first sub-case in Equation~\eqref{eq:y update}. 
It follows that the primal variables $\{x_p^*: p\in\mathcal{P}\}$ and the dual variables $\{\lambda_{u,v}^*:(u,v)\in E\}, \{\mu_{u,v}^*:(u,v) \in E\}$ satisfy the complementary slackness conditions of the optimization problem. 

Last, the optimality condition for the primal variables on the Lagrangian defined with dual variables $\{\lambda_{u,v}^*:(u,v)\in E\}$ and $\{\mu_{u,v}^*:(u,v) \in E\}$ stipulates that 
\begin{align}
\frac{1}{\sum_{p'\in\mathcal{P}_{i_p,j_p}} x_{p'}} - \sum_{(u,v) \in p}(\lambda_{u,v}^* + \mu_{u,v}^* - \mu_{v,u}^*)
 \begin{cases*}
      =0 & if $x_p > 0$ \\
      \leq 0 & if $x_p = 0$ \\
\end{cases*}, \label{eq:lagrange opt}
\end{align}
for all $p\in \mathcal{P}$. 
However, note that for any path $p$
\begin{align}
\sum_{(u,v)\in p}(\lambda_{u,v}^* + \mu_{u,v}^* - \mu_{v,u}^*) &= \sum_{(u,v)\in p} \frac{f_{u,v}^* + f_{v,u}^*}{2} + \frac{f_{u,v}^*}{2} - \frac{f_{v,u}^*}{2} \notag \\ 
&= \sum_{(u,v) \in p} f_{u,v}^*,  
\end{align}
where the first equation above follows from our mapping for $\lambda_{u,v}^*$ and $\mu_{u,v}^*$ in Equations~\eqref{eq:lambda map},~\eqref{eq:mu map}. 
Combining this with Equation~\eqref{eq:x balance}, we see that $x_p \leftarrow x_p^*$ for all $p\in \mathcal{P}$ is a valid solution to the Equation~\eqref{eq:lagrange opt}. 
Hence we conclude that $\{x_p^*: p \in \mathcal{P} \}$ and $\{\lambda_{u,v}^*:(u,v)\in E\}$, $\{\mu_{u,v}^*:(u,v) \in E\}$ are optimal primal and dual variables, respectively, for the optimization problem. 
The equilibrium rates found by \name for the parallel network topology are optimal. 

\section{Primal-Dual Algorithm Derivation}
\label{apx: primal dual derivation}

In this section, we present a formal derivation of the decentralized algorithm for computing the optimum solution of the fluid-model LP (Eq.~\eqref{eq:LP}--\eqref{eq:LP end}). 
Consider the partial Lagrangian of the LP: 
\begin{align}
L(\mathbf{x}, \lambda, \mu) = &\sum_{i,j \in V} \sum_{p\in\mathcal{P}_{i,j}}  x_p \notag \\ 
& - \sum_{(u,v)\in E}  \lambda_{(u,v)} \left[ \sum_{\substack{p\in\mathcal{P}: \\(u,v) \in p}} x_p  + \sum_{\substack{p'\in\mathcal{P}: \\ (v,u) \in p'}} x_{p'} - \frac{c_{(u,v)}}{\Delta} \right]  \notag \\
& - \sum_{(u,v)\in E} \mu_{(u,v)} \left[ \sum_{\substack{p\in\mathcal{P}: \\ (u,v)\in p}} x_p - \sum_{\substack{p'\in\mathcal{P}: \\ (v,u)\in p'}}x_{p'} \right] \notag \\
& - \sum_{(u,v)\in E} \mu_{(v,u)} \left[ \sum_{\substack{p\in\mathcal{P}: \\ (v,u)\in p}}x_{p} - \sum_{\substack{p'\in\mathcal{P}: \\ (u,v)\in p'}} x_{p'} \right],
\end{align}
where $\mu_{(u,v)}, \mu_{(v,u)}$ are Lagrange variables corresponding to the imbalance constraints (Eq.~\eqref{eq:LP balance}) in the $u$-$v$ and $v$-$u$ directions respectively.
$\lambda_{(u,v)}$ is a Lagrange variable corresponding to the capacity constraint (Eq~\eqref{eq:LP capacity}). 
Since the $\lambda$ variable does not have a direction associated with it, to simplify notation we use $\lambda_{(v,u)}$ and  $\lambda_{(u,v)}$ interchangeably to denote $\lambda_{(u,v)}$ for channel $(u,v) \in E$.  
The partial Lagrangian can be rewritten as 
\begin{align}
L(\mathbf{x}, \lambda, \mu) = &\sum_{i,j\in V} \sum_{p \in \mathcal{P}_{i,j}} x_p \left(  1 - \sum_{(u,v) \in p } \lambda_{(u,v)} - \sum_{(u,v)\in p} \mu_{(u,v)} \right. \notag \\
&\left. + \sum_{(v,u)\in p} \mu_{(v,u)} \right) + \sum_{(u,v)\in E} \lambda_{(u,v)} \frac{c_{(u,v)}}{\Delta}. 
\end{align}
Define $z_{(u,v)} = \lambda_{(u,v)} + \mu_{(u,v)} - \mu_{(v,u)}$ to denote the price of channel $(u,v) \in E$ in the $u$-$v$ direction, and $z_p = \sum_{(u,v): (u,v) \in p}  \left(\lambda_{(u,v)} + \mu_{(u,v)} - \mu_{(v, u)} \right)$ to be the total price of a path $p$. 
The partial Lagrangian above decomposes into separate terms for rate variables for each source/destination pair $\{x_p: p\in \mathcal{P} \}$.
This suggests the following iterative primal-dual algorithm for solving the LP: 
\begin{itemize}[leftmargin=*]
\item
{\bf Primal step.} Supposing the path price of a path $p$ at time $t$ is $z_p(t)$.  
Then, each sender-receiver pair $(i,j)$ updates its rates $x_p$ on each path $p \in \mathcal{P}_{i,j}$ 
as 
\begin{align}
x_p(t+1) &= x_p(t) + \alpha (1 - z_p(t)) \label{eq:rate update} \\
x_p(t+1) &= \mathrm{Proj}_{\chi_{i,j}}(x_p(t+1)),  \label{eq: projection}
\end{align}
where $\mathrm{Proj}$ is a projection operation on to the convex set $\{ x_p: \sum_{p: p \in \mathcal{P}_{i,j}} x_p \leq d_{i,j}, x_p \geq  0 \forall p \}$, to ensure the rates are feasible. 

\item
{\bf Dual step.} 
Similarly, for the dual step let $x_p(t)$ denote the flow rate along path $p$ at time $t$ and 
\begin{align}
w_{(u,v)}(t) &= \sum_{\substack{p\in\mathcal{P}: \\(u,v) \in p}} x_p(t) + \sum_{\substack{p'\in\mathcal{P}: \\ (v,u) \in p'}} x_{p'}(t)  - \frac{c_{(u,v)}}{\Delta} \label{eq: rate excess} \\
y_{(u,v)}(t) &=  \sum_{\substack{p\in\mathcal{P}: \\ (u,v)\in p}} x_p(t) - \sum_{\substack{p'\in\mathcal{P}: \\ (v,u)\in p'}}x_{p'}(t) 
\end{align}
be the slack in the capacity and balance constraints respectively for a payment channel $(u,v)$. 
Then, each channel $(u,v) \in E$ updates its prices as 
\begin{align}
\lambda_{(u,v)}(t+1) &= \left[ \lambda_{(u,v)}(t) + \eta w_{(u,v)}(t) \right]_+ \label{eq:lambda update}\\
\mu_{(u,v)}(t+1) &= \left[ \mu_{(u,v)}(t) + \kappa y_{(u,v)}(t) \right]_+ \label{eq:mu update 1} \\
\mu_{(v,u)}(t+1) &= \left[ \mu_{(v,u)}(t) - \kappa y_{(u,v)}(t) \right]_+. \label{eq:mu update 2}
\end{align}
\end{itemize}
The parameters $\alpha, \eta, \kappa$ are positive "step size" constants, that determine the rate at which the algorithm converges.  
Using standard arguments we can show that for small enough step sizes, the algorithm would converge to the optimal solution of the LP in Eq.~\eqref{eq:LP}--\eqref{eq:LP end}.

\begin{figure*}[t]
\begin{center}
\includegraphics[width=6.3in]{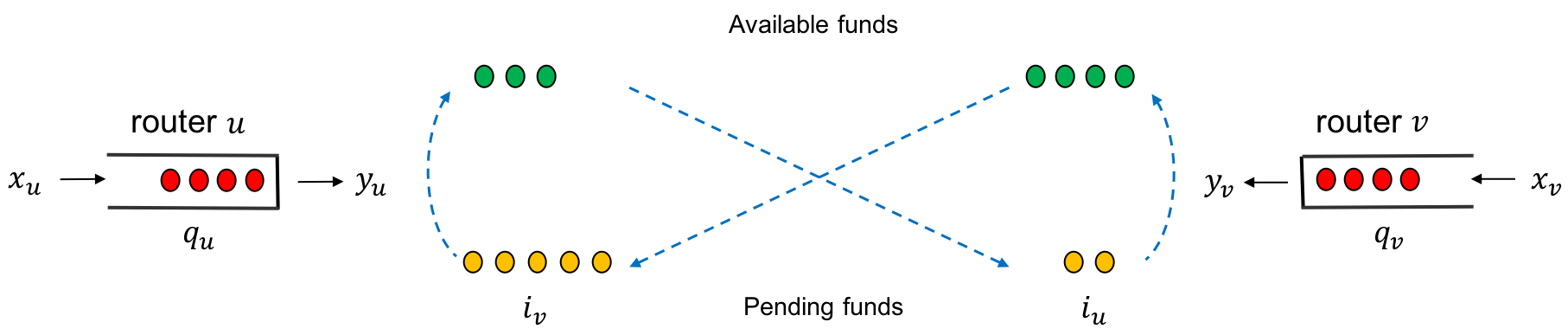}
\caption{\small Update figure. Routers queue transaction units and schedule them across the payment channel based on available capacity and transaction priorities. Funds received on a payment channel remain in a pending state until the final receiver provides the key for the hashlock.}
\label{fig:estimatingdemand}
\end{center}
\end{figure*}

The algorithm has the following intuitive interpretation. 
$\lambda_{(u,v)}$ and $\mu_{(u,v)}, \mu_{(v,u)}$ are prices that vary due to capacity constraints and imbalance at the payment channels. 
In Eq.~\eqref{eq:lambda update}, $\lambda_{(u,v)}$ would increase if the total rate on channel $(u,v)$ (in both directions) exceeds its capacity, and would decrease to 0 if there is excess capacity. 
Similarly, $\mu_{(u,v)}$ would increase (resp. decrease) if the net rate in the $u$-$v$ direction is greater (resp. less) than the net rate in the $v$-$u$ direction (Eq.~\eqref{eq:mu update 1},~\eqref{eq:mu update 2}).   
As the prices vary, an end-host with a flow on path $p$ would react according to Eq.~\eqref{eq:rate update} by increasing its sending rate $x_p$ if the total price of the path $p$ is cheap, and decreasing the rate otherwise. 
The net effect is the convergence of the rate and price variables to values such that the overall throughput of the network is maximized. 
We remark that the objective of our optimization problem in Eq.~\eqref{eq:LP} can be modified to also ensure fairness in routing, by associating an appropriate utility function with each sender-receiver pair~\cite{kelly1998rate}. 
A decentralized algorithm for such a case may be derived analogously as our proposed solution.

\section{Estimating the Demand-Capacity Gap at the Routers}
\label{apx: demand estimation}

In this section, we explain how \TheSystem estimates the total amount of demand on a channel at any time, for updating the capacity price $\lambda$ in Eq.~\eqref{eq:lambda update sys}. 
From the description of the primal-dual algorithm for the fluid model in \App{apx: primal dual derivation}, we see that updating $\lambda_{(u,v)}$ at a channel $(u,v)\in E$ requires estimating 
\begin{align}
\sum_{\substack{p\in\mathcal{P}: \\(u,v) \in p}} x_p(t) + \sum_{\substack{p'\in\mathcal{P}: \\ (v,u) \in p'}} x_{p'}(t)  - \frac{c_{(u,v)}}{\Delta} \label{eq: lambda update grad}
\end{align}
at the channel (Eq.~\eqref{eq:lambda update}). 
While the total rate at which transactions are arriving at $u$ ($\sum_{p\in\mathcal{P}: (u,v) \in p} x_p(t)$) and at $v$ ($ \sum_{p'\in\mathcal{P}:  (v,u) \in p'} x_{p'}(t)$) are straightforward to estimate, estimating $\Delta$---the average time taken for transactions to reach their destination from the channel, and for their hashlock keys to arrive at the channel---is difficult. 
In \TheSystem, we overcome this problem by estimating the quantity 
\begin{align}
\sum_{\substack{p\in\mathcal{P}: \\(u,v) \in p}} x_p(t) \Delta + \sum_{\substack{p'\in\mathcal{P}: \\ (v,u) \in p'}} x_{p'}(t) \Delta  - c_{(u,v)}, \label{eq: rearrange lambda grad}
\end{align}
instead of trying to estimate the expression in Eq.~\eqref{eq: lambda update grad}. 
Eq.~\eqref{eq: rearrange lambda grad} is simply a scaling of Eq.~\eqref{eq: lambda update grad}, but can be estimated without having to first estimate $\Delta$. 
To see this, let $\tilde{x}_u(t) = \sum_{p\in\mathcal{P}: (u,v) \in p} x_p(t)$ and $\tilde{x}_v(t) = \sum_{p'\in\mathcal{P}:  (v,u) \in p'} x_{p'}(t)$ denote the rate of transaction arrival at $u$ and $v$ respectively. 
Similarly, let $\tilde{y}_u(t)$ and $\tilde{y}_v(t)$ be the rate at which transactions are serviced from the queue at each of the routers (see Fig.~\ref{fig:estimatingdemand} for an illustration).
Eq.~\eqref{eq: rearrange lambda grad} can now be rewritten as $\tilde{x}_u(t) \Delta + \tilde{x}_v(t) \Delta  - c_{(u,v)}$ 
\begin{align}
& = \left( \frac{\tilde{x}_u(t)}{\tilde{y}_u(t)} \right) \tilde{y}_u(t) \Delta + \left( \frac{\tilde{x}_v(t)}{\tilde{y}_v(t)} \right) \tilde{y}_v(t)  \Delta  - c_{(u,v)} \label{eq: lambda grad multiply and divide} \\ 
& = \left( \frac{\tilde{x}_u(t)}{\tilde{y}_u(t)} \right) i_u(t) + \left( \frac{\tilde{x}_v(t)}{\tilde{y}_v(t)} \right) i_v(t) - c_{(u,v)},  \label{eq: lambda grad multiply}
\end{align}
where $i_u(t)$ and $i_v(t)$ are the amount of funds that are currently locked at routers $v$ and $u$ respectively (Fig.~\ref{fig:estimatingdemand}).
Since the funds used when servicing transactions at router $u$ require $\Delta$ seconds on average to become available at $v$, by Little's law the product of the average service rate $\tilde{y}_u(t)$ and average delay $\Delta$ is equal to the average amount of pending transactions $i_u(t)$ at $v$.  
Thus, Eq.~\eqref{eq: lambda grad multiply} follows from Eq.~\eqref{eq: lambda grad multiply and divide}. 
However, each of the terms in Eq.~\eqref{eq: lambda grad multiply}---the transaction arrival rates $\tilde{x}_u(t), \tilde{x}_v(t)$, service rates $\tilde{y}_u(t), \tilde{y}_v(t)$, amount of pending transactions $i_u(t), i_v(t)$---can now be readily estimated at the channel. 

Intuitively, since $i_u(t)$ is the amount of pending funds at router $v$ when transactions are being serviced at a rate $\tilde{y}_u(t)$, $\tilde{x}_u(t)i_u(t)/\tilde{y}_u(t)$ is an estimate of the amount of transactions that will be pending if transactions were serviced at a rate $\tilde{x}_u(t)$. 
As the total amount of pending transactions in the channel cannot exceed the total amount of funds escrowed $c_{(u,v)}$, the difference $\tilde{x}_u(t)i_u(t)/\tilde{y}_u(t) + \tilde{x}_v(t)i_v(t)/\tilde{y}_v(t) - c_{(u,v)}$ is exactly the additional amount of funds required in the channel to support the current rates of transaction arrival. 
Denoting $\tilde{x}_u(t)i_u(t)/\tilde{y}_u(t)$ as $m_u(t)$ and $\tilde{x}_v(t)i_v(t)/\tilde{y}_v(t)$ as $m_v(t)$, the equation for updating $\lambda$ at the routers can be written as 
\begin{align}
\lambda_{(u,v)}(t+1) = \left[ \lambda_{(u,v)}(t) + \eta \left( m_u(t) + m_v(t) - c_{(u,v)} \right. \right. \notag \\
\left. \left. + \beta \min(q_u(t), q_v(t)) \right) \right]_+,  
\end{align}
where the $\beta \min(q_u(t), q_v(t))$ term has been included to ensure the queue sizes are small as discussed in \S\ref{sec:transport layer endhosts}.

\eat{
    \section{Evaluation Datasets}
\label{app:dataset}
\begin{figure}
    \centering
    \begin{subfigure}[t]{0.48\columnwidth}
    	\includegraphics[width=\columnwidth]{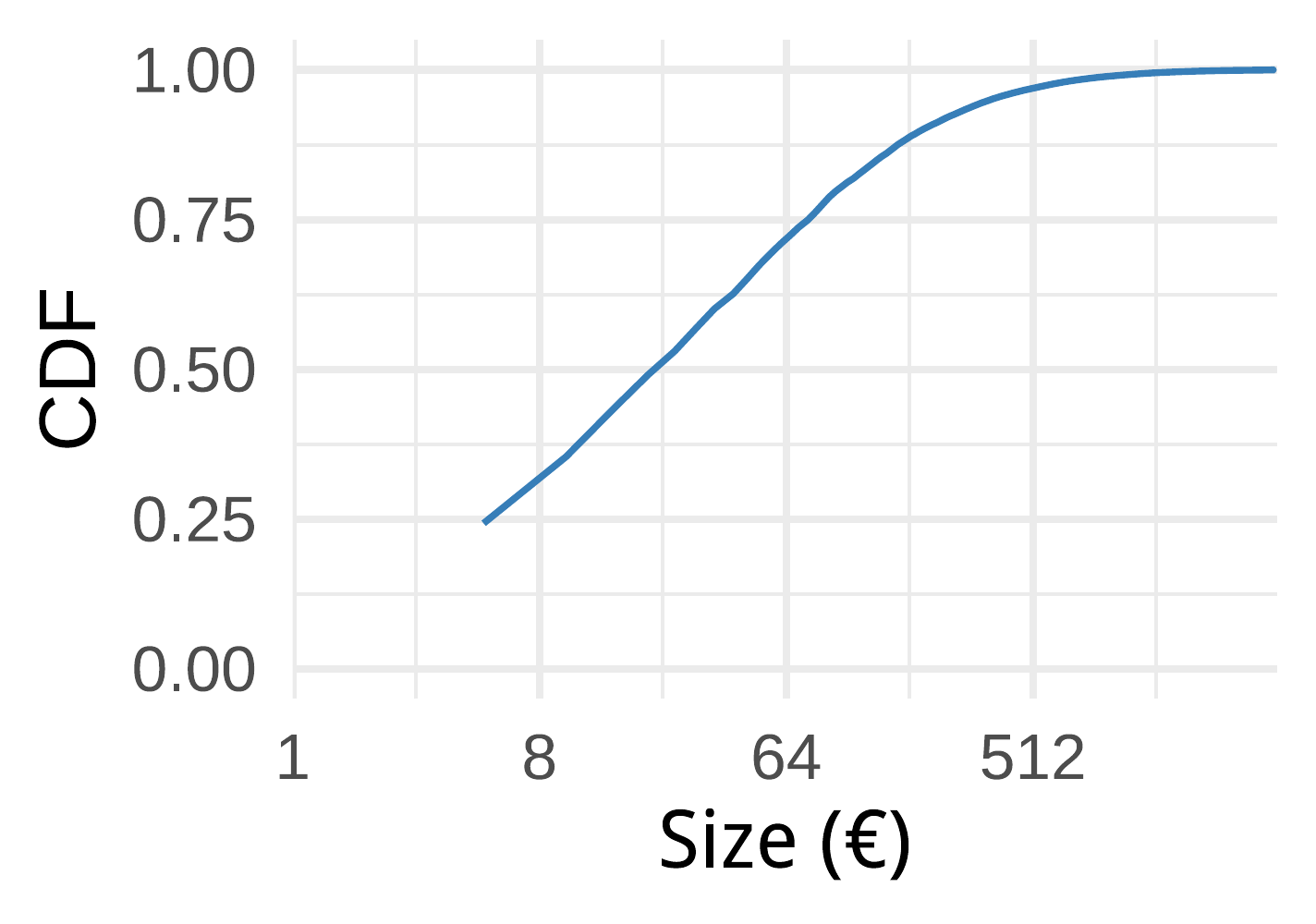}
        \caption{\small Transaction Size Distribution}
   	\label{fig:kaggle}
    \end{subfigure}
    \begin{subfigure}[t]{0.48\columnwidth}
    	\includegraphics[width=\columnwidth]{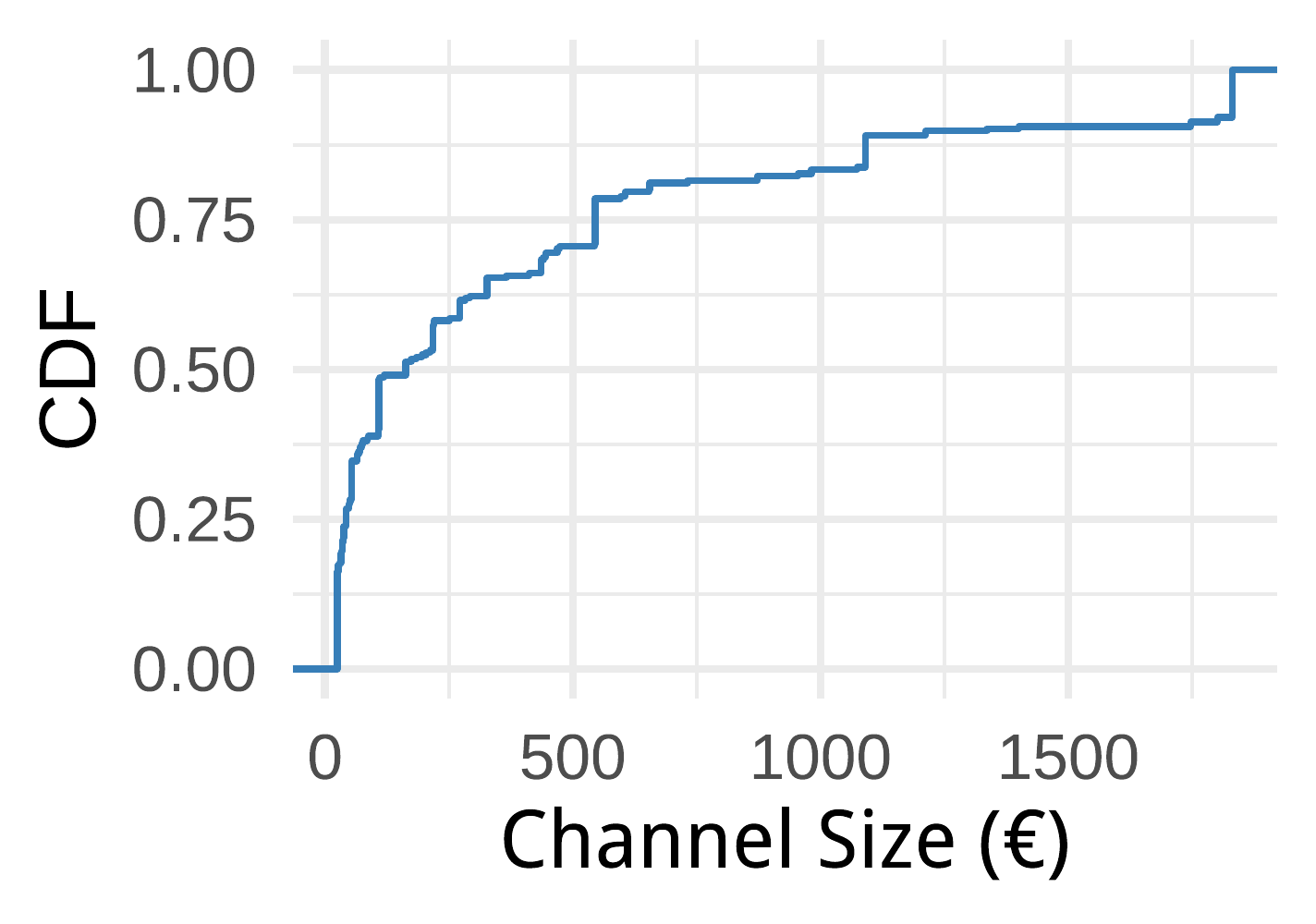}
        \caption{\small LN Channel Size Distribution}
   	\label{fig:lnd:topo:capacity}
    \end{subfigure}~
		\caption{\small Transaction dataset and channel size distribution used for
                real-world evaluations.}
    \label{fig:dataset}
\end{figure}

\Fig{dataset} shows the distribution of transaction sizes and channel sizes that were used for the evaluations
described in \Sec{sec:eval}. The transaction size distribution is drawn from credit card transaction data \cite{kaggledata} (\Fig{kaggle}), 
and has a mean of 88\eu, with
the largest transaction being 3930\eu.

To experiment with a realistic network, we set up an LND node~\cite{lightning-implementation}, and snowball sample~\cite{snowball-sampling} from the Lightning Network topology on July 15, 2019. 
The resulting \PCN has 106 nodes and 265 payment channels. 
For compatibility with our transaction dataset, we convert LND payment channel sizes from Satoshis to \eu, and cap the minimum channel size to the median transaction size of 25\eu.
Further, since the median transaction size is 25\eu, we set the bottom 15\% of the sampled subgraph's  
payment channel sizes to 25\eu, making the minimum payment channel size 25\eu. 
The resulting distribution of channel sizes (shown in \Fig{lnd:topo:capacity}) has a mean and median size of 421\eu and 163\eu respectively. We refer to this distribution as the Lightning Channel Size Distribution (LCSD)
in \Sec{sec:eval}.
We draw the per-hop RTT from from the ping times we observed from our LND node to all reachable nodes
in the Lightning Network. This results in transaction completion times (absent of queueing delay) across 7-8 hops of about  a second. 
}

\begin{figure}[t]
    \centering
    \includegraphics[width=0.6\columnwidth]{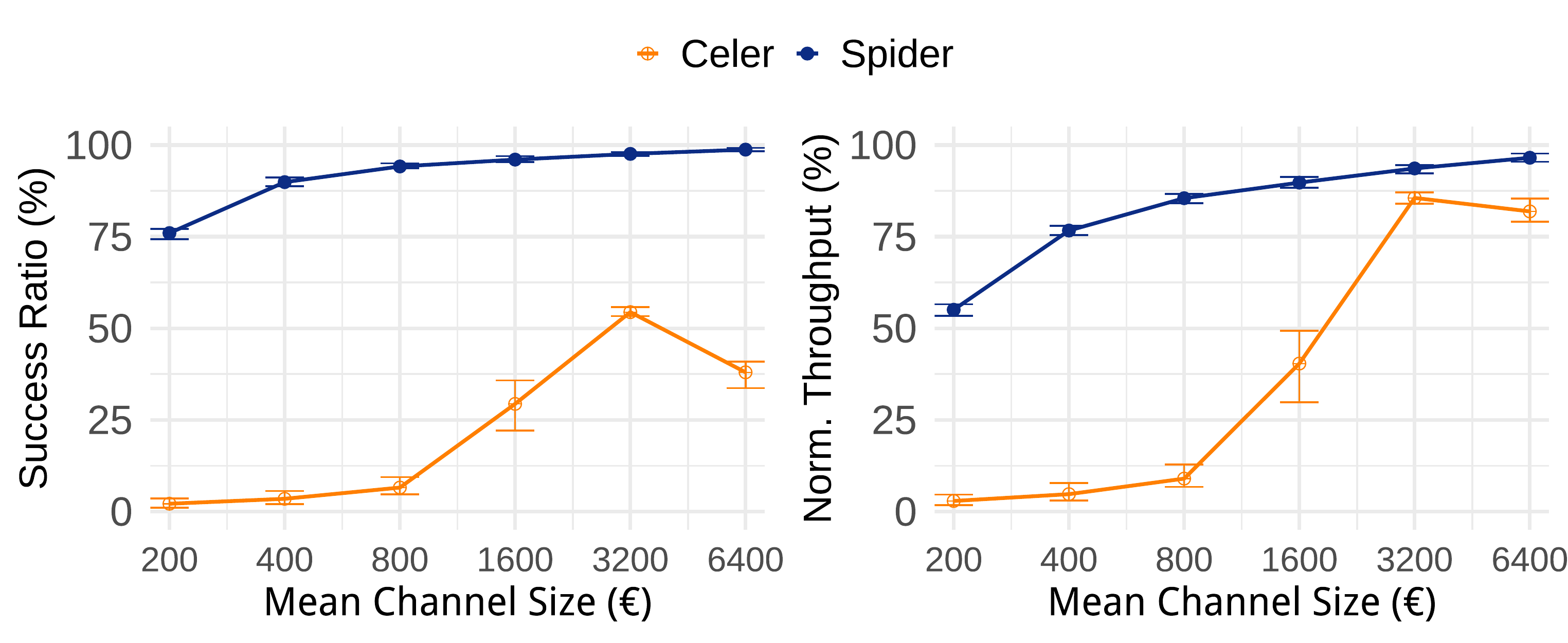}
    \caption{\small \name's performance relative to Celer on a 10 node scale free topology. \name achieves a 2x improvement
    in \SR even at Celer's peak performance. Celer's performance dips after a peak since it maintains
    larger queues at higher capacities, eventually causing timeouts.}
    \label{fig:celer}
\end{figure}

\section{Additional Results}
\subsection{Comparison with Celer}
\label{app:celer}
    We run five circulation traffic matrices for 610s on a scale free topology with 10 nodes and 25 edges 
to compare \name to Celer \cite{celer}, a back-pressure based routing scheme. Each node sends 30 txns/s
and we vary the mean channel size from 200\eu to 6400 \eu. We measure the average \SR and \SV for transactions
in the 400-600s interval and observe that \name outperforms Celer at all channel sizes.
Celer splits transactions into \tus at the source but does not source-route individual \tus. Instead, \tus for a destination
are queued at individual routers and forwarded on the link with the maximum queue and imbalance gradient for that destination. 
This approach tries to maximize \tus in queues to improve network utilization.
However, queued-up and in-flight units in PCNs hold up tokens in other parts of the network while they are in-flight
waiting for acknowledgements, reducing
its capacity. Celer transactions also use long paths, sometimes upto 18 edges in this network with 25 edges.
Consequently, tokens in Celer spend few seconds in-flight in contrast to the hundreds of milliseconds
with \name. The time tokens spent in-flight also increases with channel size since Celer tries to maintain larger queues.
Celer's performance dips once the in-flight time has increased to the point where transactions start timing out before
they can be completed. Due to computational constraints associated with large queues, we do not run Celer on larger topologies.

\begin{figure*}[t]
    \centering
    \includegraphics[width=\textwidth]{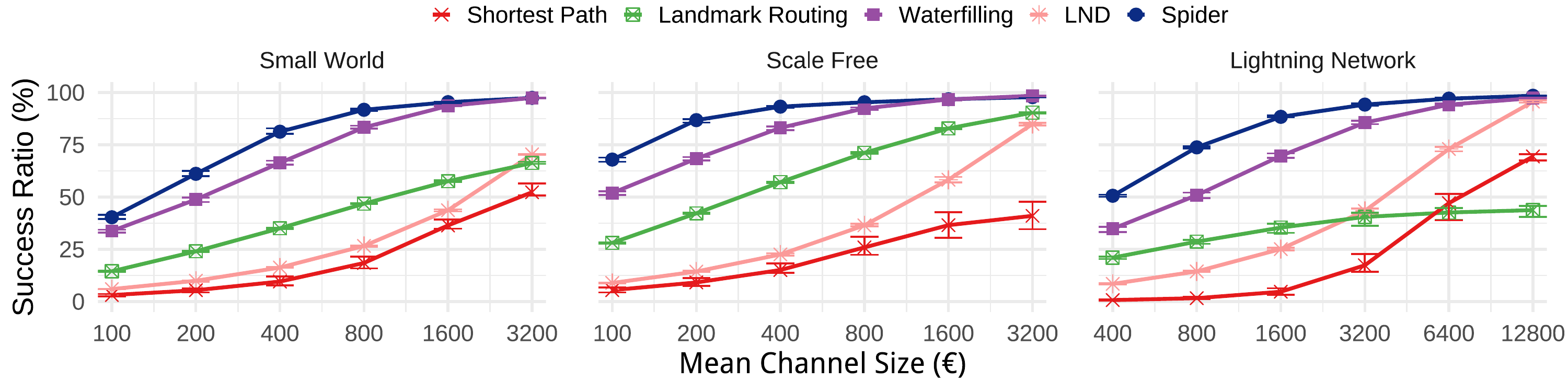}
    \caption{\small Performance of different algorithms on different topologies with equal
        channel sizes with different 
    per sender transaction arrival rates. \name consistently outperforms all other schems achieving near 100\%
    average \SR. Error-bars denote the maximum and minimum \SR across five runs. Note the log scale of the x-axes.}
    \label{fig:uniform bigexp throughput}
\end{figure*}
\begin{figure*}[t]
    \centering
    \includegraphics[width=\textwidth]{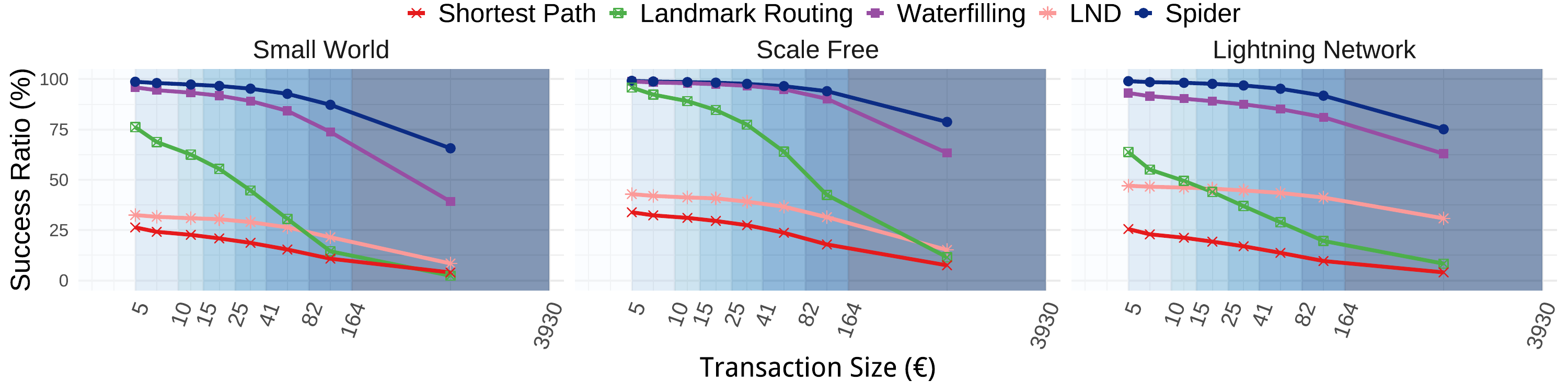}
    \caption{\small Breakdown of performance of different schemes by size of transactions completed. 
        Each point reports the \SR for transactions whose size
        belongs to the interval denoted by the shaded region. 
        Each interval corresponds roughly to 12.5\% of the CDF denoted in \Fig{kaggle}. The graphs correspond to the (right) midpoints of the corresponding Lightning sampled channel sizes in \Fig{bigexp throughput}.}
    \label{fig:uniform bigexp size}
\end{figure*}

\subsection{Circulations on Synthetic Topologies}
\label{app:uniform csd}
We run five circulation traffic matrices for 1010s on our three topologies
with all channels having exactly the tokens denoted by the channel size.
\Fig{uniform bigexp throughput} shows that across all topologies, \name outperforms the state-of-the-art schemes
on \SR.
\name is able to successfully route more than 95\% of the 
transactions with less than 25\% of the capacity required by LND. Further \Fig{uniform bigexp size} shows
that \name completes nearly 50\% more of the largest 12.5\% of the transactions attempted in the \PCN
across all three topologies. Even the waterfilling heuristic outperforms LND by 15-20\% depending
on the topology.

\begin{figure}[t]
    \centering
    \includegraphics[width=0.8\columnwidth]{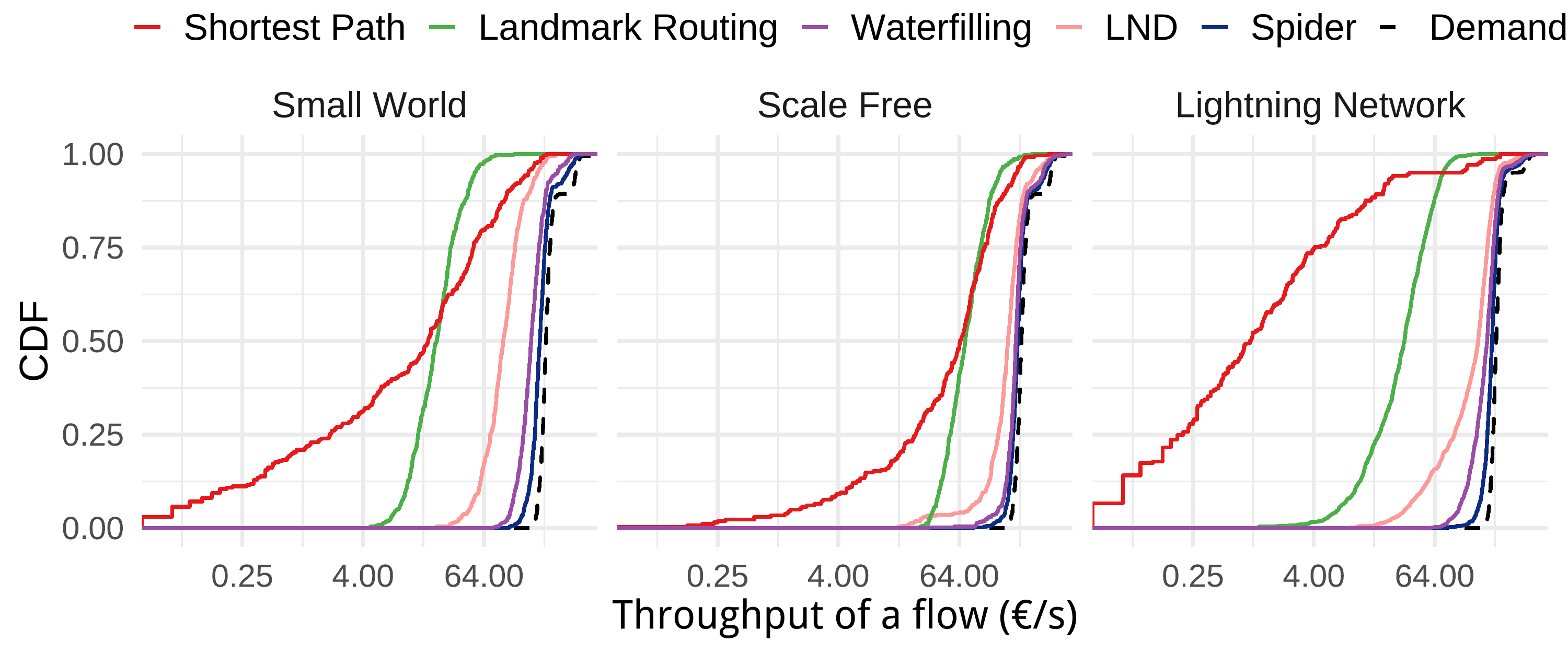}
    \caption{\small CDF of normalized throughput achieved by different flows under different schemes across topologies. \name achieves close to 100\% throughput given its proximity to the black demand line.
    \name is more vertical line than LND because it is fairer: it doesn't hurt the throughput of smaller flows to attain good overall throughput.}
    \label{fig:fairness}
\end{figure}

\subsection{Fairness of Schemes}
\label{app:fairness}
In \Sec{sec:circulation}, we show that \name outperforms state-of-the art schemes on the \SR achieved for a given 
channel capacity. Here, we break down the \SV by flows (sender-receiver pairs) to understand the fairness of the scheme
to different pairs of nodes transacting on the \PCN.
\Fig{fairness} shows a CDF of the absolute throughput in \eu/s achieved by different protocols on a single
circulation demand matrix when each sender sends an average of 30 tx/s.
The mean channel sizes for the synthetic topologies and the real
topologies with LCSD channel sizes are 4000\eu and 16880\eu respectively. We run each protocol
for 1010s and measure the \SV for transactions arriving between 800-1000s.
We make two observations: (a) \name achieves close
to 100\% throughput in all three scenarios,
(b)\name is fairer to small flows (most vertical line) and doesn't hurt the smallest flows just to benefit
on throughput. This is not as true for LND.

\label{app:dag}
\begin{figure}[h]
    \centering
    \includegraphics[width=0.5\columnwidth]{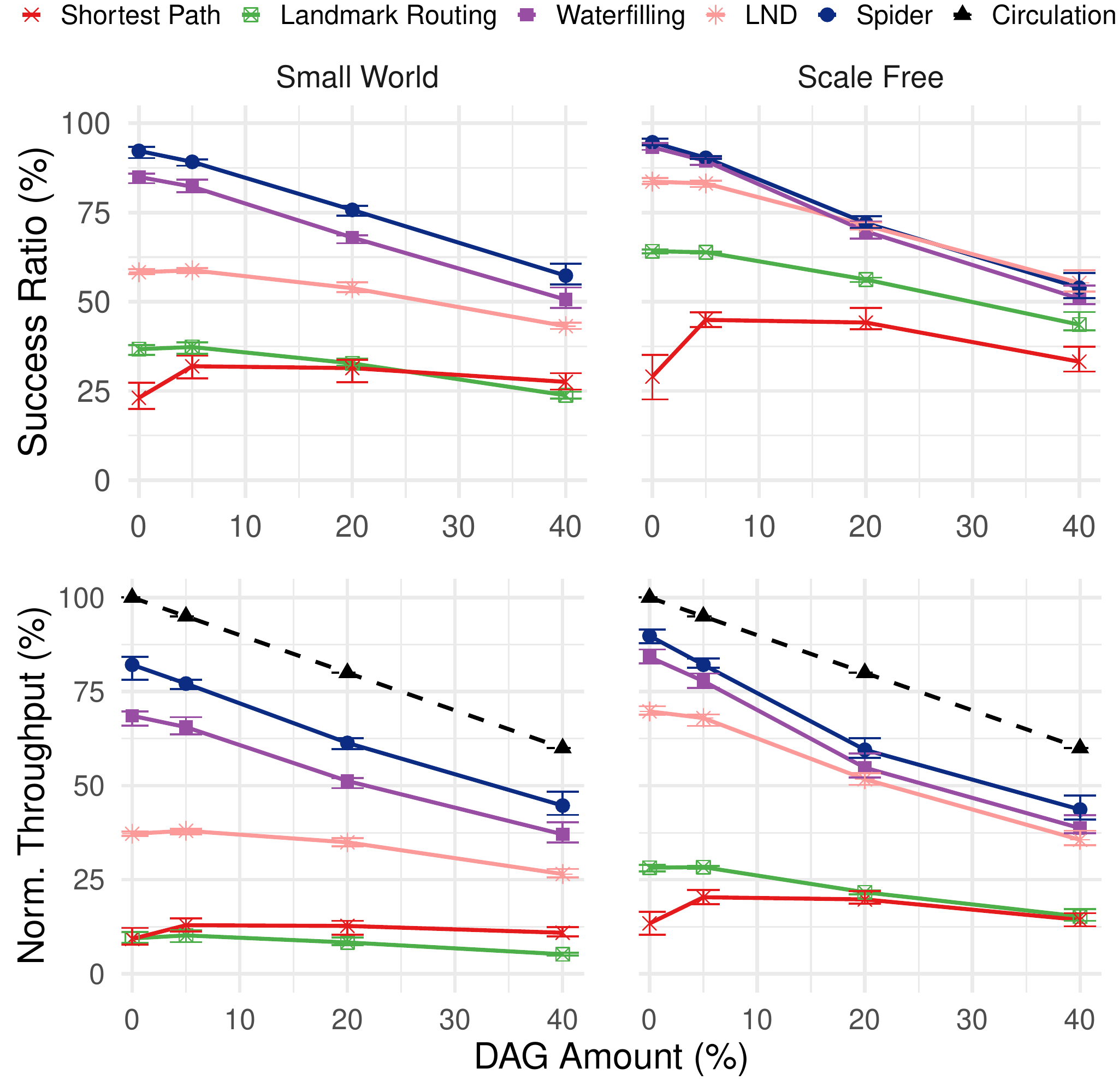}
     \caption{\small 
         Performance of different algorithms across all topologies as the DAG component in the transaction
     demand matrix is varied.  As the DAG amount is increased, the \NT achieved is further away 
     from the expected optimal circulation throughput. The gap is more pronounced on the real topology.}
    \label{fig:synthetic topo dag throughputs}
\end{figure}

\begin{figure}[t!]
    \centering
    \includegraphics[width=0.5\columnwidth]{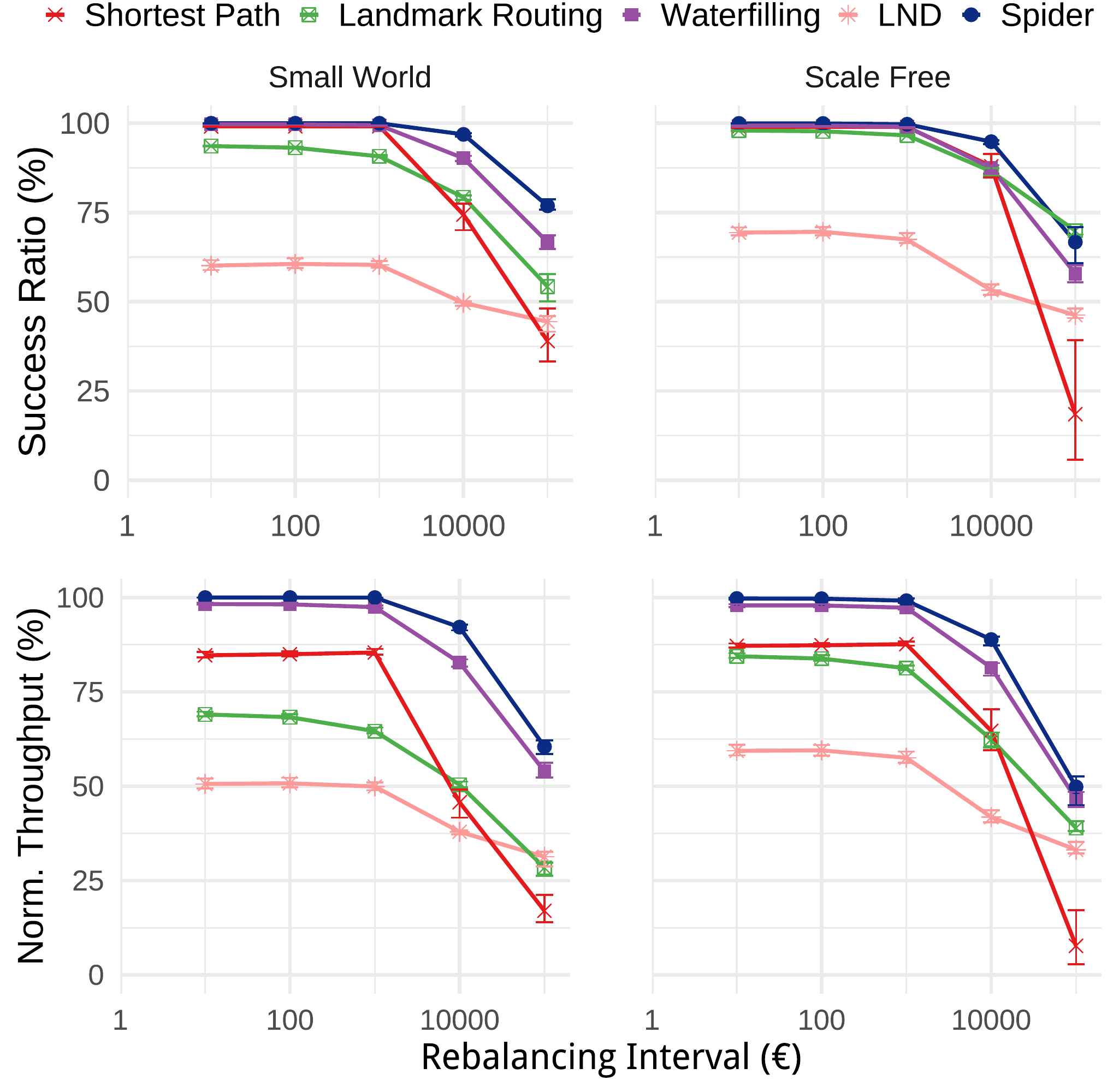}
     \caption{\small 
         Performance of different algorithms across all topologies as the DAG component in the transaction
     demand matrix is varied.  As the DAG amount is increased, the \NT achieved is further away 
     from the expected optimal circulation throughput. The gap is more pronounced on the real topology.}
    \label{fig:synthetic topo rebalancing throughputs}
\end{figure}

\subsection{DAG Workload on Synthetic Topologies} 
\Fig{synthetic topo dag throughputs} shows the effect of adding a DAG component to the transaction demand
matrix on the synthetic small world and scale free topologies. We observe the \SR and \NT of different schemes
with five different traffic matrices with 30 transactions per second per
sender under 5\%, 20\%, 40\% DAG components respectively.
No scheme is able to achieve the maximum throughput. However, the achieved throughput is closer to the maximum 
when there is a smaller component of DAG in the demand matrix. 
This suggests again that the DAG affect PCN balances in a way that also prevents the circulation from going through. 
We investigate what could have caused this and how pro-active on-chain rebalancing could alleviate this in 
\Sec{sec:dag}.

\Fig{synthetic topo rebalancing throughputs} shows the 
\SR and \NT achieved by different schemes when rebalancing is enabled for the 20\% DAG traffic
demand from \Fig{synthetic topo dag throughputs}. \name is able to achieve over 95\% \SR  and 90\% \NT 
even when its routers balance 
only every 10,000 \eu while LND is never able to sustain more than 75\% \SR even when rebalancing for every 10\eu
routed. This implies that \name makes PCNs more economically 
viable for both routers locking up funds in payment channels
and end-users routing via them
since they need far fewer on-chain rebalancing events to sustain high throughput and earn routing fees.


\end{document}